\documentclass[%
 reprint,
 superscriptaddress,
 amsmath,
 amssymb,
 aps,
 pra,
]{revtex4-2}

\usepackage{graphicx}
\usepackage{dcolumn}
\usepackage{bm}

\usepackage{float}
\usepackage{blindtext}
\usepackage{amsmath,amsfonts,amsthm,physics}
\usepackage{comment}
\usepackage{lipsum}
\usepackage[export]{adjustbox}
\usepackage{xprintlen}
\usepackage{xcolor}
\usepackage{placeins}
\usepackage{hyperref}

\newcommand{\LL}{\left}
\newcommand{\RR}{\right}
\newcommand{\id}{\mathbb{I}}
\newcommand{\sz}{\sigma_Z}

\newcommand{\e}{\varepsilon}

\renewcommand{\selectlanguage}[1]{}

\begin{document}

\preprint{APS/123-QED}

\title{Co-transmission of classical data and continuous-variable entanglement over a single quantum channel}

\author{Nicholas Zaunders} \email{n.zaunders@uq.edu.au}
\affiliation{
    School of Mathematics and Physics, University of Queensland, St Lucia, Queensland 4072, Australia.
}

\author{Timothy C. Ralph}
\affiliation{
    School of Mathematics and Physics, University of Queensland, St Lucia, Queensland 4072, Australia.
}

\begin{abstract}
Displacement-based simultaneous quantum-classical communications (SQCC) protocols, as originally proposed, are generally incompatible with the majority of useful quantum communication schemes, such as entanglement distribution or repeater-based quantum key distribution: direct measurement of the classical signal also measures and destroys the quantum state, leaving point-to-point Gaussian quantum key distribution as the only quantum communication scheme amenable to integration with SQCC. In this work, we apply the classically-modulated quantum communication protocol proposed by Zaunders and Ralph \cite{zaunders_generalised_2026} to circumvent this issue and demonstrate the distribution of continuous-variable Gaussian entanglement simultaneously with classical information. We characterise the quality of the distributed entangled state and outline how the scheme is suitable for use in repeater-based networks. Lastly, we compute the secret key generation rate of the Gaussian CMQC scheme in the point-to-point case and compare it to the equivalent point-to-point SQCC protocol.
\end{abstract}

\maketitle

\section{Introduction} \label{sec:intro}
The cornerstone of modern quantum technology is the generation and harnessing of entanglement, a uniquely non-classical property of multipartite quantum states which acts as a kind of resource for performing useful functions in quantum information processing \cite{nielsen_quantum_2010}. Entanglement shared between two or more physically distant locations, in particular, can be exploited to achieve outcomes which would ordinarily be entirely inaccessible to purely-classical communication protocols, including quantum key distribution \cite{bennett_quantum_2014, ekert_quantum_1991, gisin_quantum_2002, ralph_continuous_1999, grosshans_quantum_2003}, quantum teleportation \cite{bennett_teleporting_1993, bennett_purification_1996, braunstein_criteria_2000}, clock synchronisation and navigation \cite{jozsa_quantum_2000, yurtsever_lorentz-invariant_2002, komar_quantum_2014, ilo-okeke_remote_2018}, distributed quantum computing \cite{cirac_distributed_1999} and quantum sensing \cite{gottesman_longer-baseline_2012, degen_quantum_2017, zhuang_distributed_2018, xia_repeater-enhanced_2019}. For this reason a chief aim of quantum research is the development of entanglement distribution schemes, which reliably and efficiently generate physically-separated entangled quantum modes via the transmission of entangled states.

One approach to entanglement distribution is to use continuous-variable (CV) optical entangled states. Continuous-variable states utilise an infinite-dimensional Hilbert space, where it is most sensible to characterise measurement outcomes in terms of their distribution in a continuous phase space composed of the quadratures of the electromagnetic field \cite{walls_quantum_2008,gerry_introductory_2004}, as opposed to discrete-variable states, which exist in a finite Hilbert space. An advantage of continuous-variable resource states is that continuous-variable states can be generated reliably, stably and deterministically and do not require heralding or specialised laboratory equipment to prepare or measure \cite{bachor_guide_2019}. These properties make continuous-variable quantum information protocols attractive for practical deployment in future quantum networks \cite{usenko_continuous-variable_2025}.

Another advantage of continuous-variable protocols is that the additional degrees of freedom inherent to the larger Hilbert space allows classical information to be encoded simultaneously in a practical way. In the case of Gaussian-modulated coherent-state quantum key distribution (QKD), the encoding is performed by modulating the displacement of the quantum state in the phase space \cite{qi_simultaneous_2016, qi_noise_2018} and was termed simultaneous quantum-classical communication (SQCC); an experimental implementation of the SQCC-QKD protocol was shown in \cite{kumar_experimental_2019}. Later works \cite{zaunders_enhanced_2025} presented a formalised version of the SQCC scheme and demonstrated composable security under general attacks.

A primary limitation of the existing SQCC scheme is that it is strictly limited to key distribution protocols, since it is assumed that decoupling of the quantum and classical data streams is performed in postprocessing, i.e. after measurement of the hybrid quantum-classical state has taken place. In this work, we circumvent this issue by utilising the classically-modulated quantum communication (CMQC) protocol proposed by the current authors \cite{zaunders_generalised_2026} to analyse the specific case of the simultaneous distribution of classical information and continuous-variable entanglement under practical requirements for classical performance. Such a scheme has an immediate advantage over either ordinary entanglement distribution or optical-frequency classical communication, when viewed from the perspective of future quantum-integrated networks; by integrating the classical and quantum information into a single hybrid quantum-classical signal, existing classical communication networks can be retrofitted to distribute entanglement synchronously without any major loss in quality. Combined with high-lifetime optical quantum memories \cite{nguyen_quantum_2019,liu_heralded_2021,ortu_storage_2022},the Gaussian CMQC protocol could for example allow for any node within a classical network to passively generate and store high-quality entanglement with every other node in the network while maintaining classical communications uptime and without requiring substantial increases in computational or hardware cost. Alternatively, the CMQC scheme could be used to integrate classical communications natively to quantum protocols which require distributed sources of entanglement as well as LOCCs to operate (for example when entanglement swapping between nodes), allowing the protocol to be executed compactly over a single fibre line or free-space channel.

This paper is structured as follows. In Section \ref{sec:protocols}, we describe the hybrid scheme introduced in \cite{zaunders_generalised_2026}, starting from the required entangled resources and moving onto the classical modulation and transmission. We then describe the teleportation stage which decouples the classical and quantum datastream, and outline how Bob reconstructs each classical symbol via his teleportation measurements. Lastly, we describe the feedforward operations required for Bob to return the joint output state to an appropriate two-mode-squeezed entangled state. Section \ref{sec:analysis} explicitly analyses an implementation of the protocol assuming a quadrature phase-shift-keyed classical modulation and a Gaussian thermal-noise lossy channel; we define the classical encoding in terms of a discrete alphabet and signal strength and convert to a phase-space modulation of the distributed entangled signal. We characterise the state after transmission. We then perform the teleportation step and associated measurement, describe how Bob interprets his measurement results to reconstruct Alice's classical datastream, and explicitly compute the bit-error rate of the classical signal as a function of initial entanglement, classical signal strength and channel characteristics. We then characterise the feedforward operations performed by Bob to reproduce a two-mode squeezed vacuum state after teleportation and explicitly compute the output state obtained by Alice and Bob at the end of the protocol for a full classical alphabet, which includes non-Gaussian effects arising from classical bit errors. We quantify the output state in terms of its Gaussian components. In Section \ref{sec:results}, we provide a lower bound on the quantum performance of the protocol in terms of the entanglement of the output state. We provide an analysis in terms of the reverse coherent information, entanglement of formation, and asymptotic secret-key rate as a function of channel loss for realistic conditions in the regime of ideal (unphysical) teleportation as well as more realistic implementations. Section \ref{sec:discussion} then concludes with a summarisation of our findings and a discussion of some future applications of the protocol to quantum networking.

\section{Protocol description} \label{sec:protocols}
\begin{figure*}
    \centering
    \includegraphics[width = \textwidth]{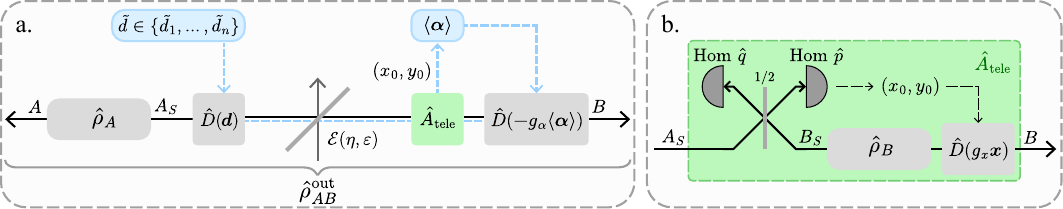}
    \caption{(a) Schematic of the Gaussian CMQC protocol. The flow of quantum and classical information is represented by black and blue dashed lines respectively. Bob and Alice each prepare a two-mode entangled state $\hat \rho_{A,B}$. Alice encodes her classical information by displacing the outgoing mode of her state by $\hat D(\bm d)$, depending on the symbol $\tilde d$ she draws from the classical alphabet $\tilde d_1, \dots, \tilde d_n$, and transmits the mode to Bob over the lossy channel $\mathcal{E}$ characterised by transmissivity $\eta$ and thermal noise $\e$. Upon reception, Bob teleports the incoming mode onto his outgoing mode by mixing Alice's signal with one arm of his entangled state on a balanced beamsplitter and performing a dual homodyne measurement, the measurement outcomes $\hat q_{A_s}, \hat p_{A_s} \longrightarrow \bm{x} = (x_0, y_0)^T$ of which are used to perform the teleportation unitary $\hat D(g_x\bm x)$ on the teleported mode $B$ (b). The outcome $\bm{x}$ is also used to generate an estimate $\langle \bm{\alpha\rangle}$ of Alice's sent classical symbol; Bob performs the displacement $\hat D(-g_\alpha\langle \bm \alpha \rangle)$ on his mode to probabilistically return the state to zero-mean. The net output of the protocol is a non-Gaussian zero-mean ensemble two-mode entangled state $\hat \rho_{AB}^\mathrm{out}$. The choice of electronic gain $g_x$ when enacting the teleportation $\hat A_\mathrm{tele}$ affects the properties of the teleporter: we assume $g_x = -\sqrt{2} \tanh r_B$ such that in the limit of many exchanged states the teleporter map $\hat \rho \longrightarrow \hat A_\mathrm{tele} \hat \rho \hat A_\mathrm{tele}^\dagger$ is equivalent to the pure-loss channel $\hat \rho \longrightarrow\Lambda[\tau](\hat \rho)$, where $\tau = \tanh^2 r_B$.}
    \label{fig:pubfig_protocol_diagram}
\end{figure*}

Based off the protocol proposed in \cite{zaunders_generalised_2026}, the structure and operating procedure of the Gaussian protocol is as follows (Fig. \ref{fig:pubfig_protocol_diagram}.a).

\textbf{State preparation and classical modulation.} To begin, Alice (Bob) generates a two-mode squeezed vacuum state $\hat \rho_{A}$ ($\hat \rho_{B}$) of squeezing $r_A$ ($r_B$). To encode the classical information, Alice performs a phase-space displacement operation on the outgoing mode of her state by an amount $ d_k \in \mathbb{R}^2$ \cite{qi_simultaneous_2016, qi_noise_2018, zaunders_enhanced_2025}. The size and direction of this displacement, which has the effect of translating the mean of the outgoing mode in the phase space, is drawn from some suitable coherent-state classical protocol with alphabet $\{\tilde d_1, \dots, \tilde d_n \}$ such that each displacement is uniquely associated with a classical symbol and corresponding binary sequence.

\textbf{Transmission.} Alice then transmits her displaced mode containing both the entanglement and classical modulation to Bob via a lossy bosonic channel $\mathcal{E}$ characterised by a transmissivity $\eta$. We also model a thermal-noise input to the channel of mean photon number $\overline{n}~=~\eta \varepsilon / 2(1 - \eta)$, such that the equivalent excess-noise contribution to the signal in shot-noise units (SNU) is $\varepsilon$.

\textbf{Demodulation}. In order to extract the classical signal, Bob must determine some way of measuring the displacement of the mode received from Alice, since this will allow him to calculate some estimator of Alice's initial displacement from which he can infer Alice's intended classical symbol. In previous proposals \cite{qi_noise_2018, zaunders_enhanced_2025} dealing with classical modulation in CV-QKD, this is easily performed via a direct measurement of the quadratures via either homodyne or heterodyne detection, with the measurement outcome containing both the classical information as well as the correlated quantum values needed to distil a secret key. However, it is clear that direct measurement is not suitable when we require the protocol to distribute entanglement, since any measurement by Bob on the received mode would collapse the state. Bob therefore needs some way to demodulate the classical and quantum signal without destroying the quantum state via measurement.

To extract the signal without directly measuring it, Bob utilises his entangled state $\hat \rho_B$ to perform a continuous-variable teleportation \cite{braunstein_quantum_2005} on the received mode. He does this by mixing one mode of his two-mode squeezed state with the incoming mode on a balanced beamsplitter and measuring the output modes of the beamsplitter via homodyne detection on $\hat q$ and $\hat p$ respectively. Crucially, his measurement values are also an estimator of the classical signal, with the homodyne measurement on $\hat q$ ($\hat p$) measuring the displacement of the signal arising from the classical scheme in the $\hat q$ ($\hat p$) direction. In this way Bob can simultaneously measure and estimate Alice's classical signal while preserving the transmitted entanglement by teleporting the correlations onto his auxiliary entangled state via a measurement-dependent displacement unitary (Fig. \ref{fig:pubfig_protocol_diagram}.b).

\textbf{Reconstruction of the quantum state.} The fact that the measurement outcomes predict Alice's classical displacement also allows Bob to additionally remove the classical displacement that is teleported onto his outgoing mode (alongside the quantum correlations) by performing the inverse displacement operation, in order to return the mode to zero-mean. Unlike the displacement arising from the teleportation, however, the removal of the classical displacement is probabilistic: there is a non-zero probability $e_S$ that Bob predicts the wrong classical symbol based on his measurements, which are necessarily imperfect. In this case, Bob would perform the wrong inverse displacement. The net result of such mistakes is to introduce a non-Gaussian noise source into his outgoing quantum mode.

\section{Analysis} \label{sec:analysis}
We now proceed to analyse the protocol, with the aim of characterising the joint entangled state $\hat \rho_{AB}^\mathrm{out}$ shared by Alice and Bob after transmission, teleportation and correction. For convenience, we consider first a simplified case where Alice sends a classical symbol with a fixed arbitrary displacement, since this allows us to show most key results without a major loss of generality. We then conclude the analysis by extending our results to the full protocol, where Alice's classical symbol is chosen randomly from a predetermined classical alphabet. A full derivation is presented in Appendix \ref{app:derivation}.

\subsection{State preparation and transmission}
We begin by assuming Alice and Bob's entangled resources are two-mode squeezed vacuum states of squeezing $r_A$ and $r_B$ respectively. These states are both Gaussian, and so are uniquely characterised by their mean vectors and covariance matrices in terms of the field quadrature operators:
\begin{align}
    \bm{\mu}_{A} = \bm{\mu}_{B} &= (0, 0, 0, 0)^T \\
    V_{AA_S} &= \begin{pmatrix}
        \cosh 2r_A \ \id & \sinh2r_A \ \sz \\
        \sinh2r_A \ \sz & \cosh2r_A \ \id
    \end{pmatrix} \\
    V_{B_SB} &= \begin{pmatrix}
        \cosh 2r_B \ \id & \sinh2r_B \ \sz \\
        \sinh2r_B \ \sz & \cosh2r_B \ \id
    \end{pmatrix}.
\end{align}
Alice and Bob retain the mode $A$ and $B$ respectively; the outgoing modes $A_S$ and $B_S$ are used for the transmission and teleportation stages. To encode each classical message, Alice displaces her outgoing mode $A_S$ according to a QPSK protocol, where her four possible classical symbols $\{ \tilde d_1, \tilde d_2, \tilde d_3, \tilde d_4 \}$ associated with the classical bitstrings $\{00, 01, 11, 10\}$ respectively are defined by
\begin{align}
    \tilde d_k &= d \exp\left( \frac{i(2k-1)\pi}{4} \right), \ k = 1,2,3,4.
\end{align}
Each symbol then corresponds to a displacement operation in the phase space $\hat D(\bm{d}_k)$, for vector displacement
\begin{align}
    \bm{d}_k &= \left( d_x, d_y \right)^T
\end{align}
where $d_x = \mathrm{Re}[\tilde d_k]$, $d_y = \mathrm{Im}[\tilde d_k]$. 

We consider initially the case where Alice only sends the symbol $00$, i.e. she displaces the outgoing mode $A_S$ of her state by the fixed amount $\hat D(\bm{d}_1)$. The covariance of her state remains unchanged, while the mean vector now becomes
\begin{align}
    \bm{\mu}_{AA_S}^{d_1} &\longrightarrow (0, 0, d_x, d_y)^T.
\end{align}
Next, Alice's displaced mode $A_S$ is transmitted to Bob through the lossy channel $\mathcal{E}$; we model this as a beamsplitter of transmittance $\eta$ and thermal-noise contribution $\e$. The effect of the channel is to transform Alice's displaced entangled state, which we label $\hat \rho_{AA_S}^{d_1}$, into a mixed Gaussian two-mode squeezed state characterised by
\begin{align}
    \bm{\mu}_{AA_S}^{d_1} &= (0, 0, \sqrt{\eta} d_x, \sqrt{\eta} d_y)^T \\
    V_{AA_S}^{d_1} &= \begin{pmatrix}
        a_{00} \id & c_{00} \sz \\
        c_{00} \sz & b_{00} \id
    \end{pmatrix}
\end{align}
where
\begin{align}
    a_{00} &= \cosh 2r_A \\
    b_{00} &= \eta(\cosh 2r_A + \e - 1) + 1 \\
    c_{00} &= \sqrt{\eta} \sinh 2r_A.
\end{align}
After transmission, the joint state $\hat \rho_{AA_SB_SB}^{d_1}$ shared by Alice and Bob is the four-mode Gaussian state with mean and covariance matrix
\begin{align}
    \bm{\mu}_{AA_SB_SB} &= (0, 0, \sqrt{\eta} d_x, \sqrt{\eta} d_y, 0, 0, 0, 0)^T \\
    V_{AA_SB_SB} &= V_{AA_S}^{d_1} \oplus V_{B_SB}.
\end{align}

\subsection{Teleportation and classical data estimation}
In order to extract the classical data stream from the combined classical-quantum signal, Bob performs a continuous-variable teleportation of Alice's mode $A_S$ onto his retained mode $B$ by first mixing the sent mode with his entangled state and measuring the output via a dual homodyne detection \cite{braunstein_quantum_2005}. This allows him to preserve the entangled state while also allowing him to estimate the classical signal via his measurement outcomes.

The first step, mixing between the signal mode $A_S$ and Bob's auxiliary entangled mode $B_S$, is done on a balanced beamsplitter, producing the four-mode entangled Gaussian state $(\hat \rho_{AA_SB_SB}^{d_1})'$ with mean vector and covariance matrix
\begin{widetext}
    \begin{align}
        \bm{\mu}_{AA_SB_SB}^{d_1}{'} &\equiv (0, 0, \alpha_x, \alpha_y, \alpha_x, \alpha_y, 0, 0)^T  \\
        V_{AA_SB_SB}^{d_1}{'} &= \begin{pmatrix}
            a_{00} \id & \frac{c_{00}}{\sqrt{2}} \sz & \frac{c_{00}}{\sqrt{2}} \sz & 0 \\
            \frac{c_{00}}{\sqrt{2}} \sz & \frac{b_{00} + \cosh2r_B}{2} \id & \frac{b_{00} - \cosh2r_B}{2} \id & -\frac{\sinh 2r_B}{\sqrt{2}} \ \sz \\
            \frac{c_{00}}{\sqrt{2}} \sz & \frac{b_{00} - \cosh 2r_B}{2} \id & \frac{b_{00} + \cosh 2r_B}{2} \id & \frac{\sinh 2r_B}{\sqrt{2}} \ \sz \\
            0 & -\frac{\sinh 2r_B}{\sqrt{2}} \ \sz & \frac{\sinh 2r_B}{\sqrt{2}} \ \sz & \cosh 2r_B \id
        \end{pmatrix} \label{eq:pre_measurement_cov}
    \end{align}
\end{widetext}
where for ease we have defined 
\begin{align}
    \bm{\alpha}_1 &= \sqrt{\eta} \bm{d}_1 / \sqrt{2} \equiv (\alpha_x, \alpha_y).
\end{align}
Bob then performs a dual homodyne measurement on the two output modes of the beamsplitter, measuring the $\hat q$ quadrature of the mode $B_S$ and obtaining the measurement result $x_0$, and measuring the $\hat p$ quadrature of the mode $A_S$ and obtaining the measurement result $y_0$ respectively. This is equivalent to projection onto the quadrature eigenstates corresponding to the correct measurement outcomes, which has projector
\begin{align}
    \hat \Pi &\equiv \id \otimes \ket{\hat p = y_0}\bra{\hat p = y_0}_{A_S} \otimes \ket{\hat q = x_0}\bra{\hat q = x_0}_{B_S} \otimes \id.
\end{align}
The probability $P^{d_1}(x_0, y_0)$ of obtaining a specific measurement result $\bm{x} = (x_0, y_0)^T$ is given by the norm of the state post-measurement, conditional on obtaining the measurement $\bm{x}$:
\begin{align}
    P^{d_1}(x_0, y_0) &= \frac{1}{\pi}\frac{\exp \left(-\frac{1}{2}\frac{(\sqrt{2}x_0-\alpha_x)^2+(\sqrt{2}y_0-\alpha_y)^2}{b_{00} + \cosh (2 r_B)}\right)}{b_{00}+\cosh(2r_B)}.
\end{align}
The simplest way for Bob to reconstruct the classical signal from his measurements is to note that, since the measurement outcome $x$ is a random variable with bivariate Gaussian distribution $P^{d_1}(x_0,y_0)$ centred on $\bm{\alpha}_1$, then with high probability the sign of the outcomes $x_0,y_0$ predict the sign of the received classical displacement and consequently Alice's sent symbol: if $x_0 \geq 0$, then it is most likely that $\alpha_x \geq 0$ and so the first bit of the symbol is $0$; if $y_0 \geq 0$, then it is most likely that $\alpha_y \geq 0$ and so the second bit of the symbol is $0$, and so on. Equivalently, obtaining $x_0 \geq 0$, $y_0\geq 0$ simultaneously implies the classical symbol $\tilde d_1$ or $00$, etc. 

The bit-error rate for bits encoded in this fashion is by definition equal to the probability that Bob obtains a measurement result predicting a binary $1$, i.e. $x < 0$, when the sent classical symbol encodes binary $0$, i.e. for $\alpha_x \geq 0$, or vice versa:
\begin{align}
    e_C &= P(x_0 < 0, y_0 \ | \ \alpha_x \geq 0) \\
    &\equiv \int \int_0^\infty \dd x_0 \ \dd y_0 \ P^{d_1}(x_0, y_0) \\
    &= \frac{1}{2} \mathrm{erfc}\left( \frac{1}{\sqrt{2}} \frac{\alpha_x}{\sqrt{b_{00} + \cosh 2r_B}} \right).
\end{align}
It is not difficult to see that by the symmetry of $P^{d_1}$ and the classical encoding the error rate for bits encoded on the $\hat p$ quadrature is also $e_C$, and in general the bit-error rate for either quadrature for any symbol is $e_C$. It thus follows that the bit-error rate of the whole protocol is $e_C$.

\subsection{Reconstructing the entangled state}
Lastly, Bob applies the correct feed-forward operations on the output mode of his teleportation, based on his measurement results, to obtain a zero-mean two-mode squeezed Gaussian state.

To propagate the state, Bob performs the unitary displacement $\hat D(g_x \bm{x})$ on his mode. By choosing the gain parameter $g_x$ such that $g_x = -\sqrt{2}\tanh r_B$ the teleporter is converted into an effective pure-loss channel \cite{ralph_continuous_1999} with transmissivity $\tau = \tanh^2r_B$; in the ensemble limit (i.e. in the limit of many shots, where each moment is averaged over the distribution of the measurement results $x_0, y_0$) the mean of the state approaches 
\begin{align}
    \bm{\mu}_B^{d_1} \longrightarrow \sqrt{\tau} \bm{\alpha_1}.
\end{align}
Removing the remaining classical displacement is less straightforward. Unlike the measurement outcomes $\bm{x}$, Bob can only estimate the classical displacement probabilistically and so cannot directly return the state to zero-mean. Instead, Bob performs an approximate displacement operation $\hat D(-g_\alpha\langle \bm{\alpha} \rangle)$ such that
\begin{align}
    \hat D(-g_\alpha\langle \bm{\alpha} \rangle) &= \begin{cases}
        \hat D(-g_\alpha \bm{\alpha}_1) & x_0 \geq 0, y_0 \geq 0 \\
        \hat D(-g_\alpha \bm{\alpha}_2) & x_0 \geq 0, y_0 < 0 \\
        \hat D(-g_\alpha \bm{\alpha}_3) & x_0 < 0, y_0 \geq 0 \\
        \hat D(-g_\alpha \bm{\alpha}_4) & x_0 < 0, y_0 < 0 \\
    \end{cases}
\end{align}
for $\bm{\alpha}_k = \sqrt{\eta}\bm{d}_k/\sqrt{2}$ and $g_\alpha = \sqrt{\tau}$. The net effect of these two operations is thus to enact the mean transformation
\begin{align}
    \bm{\mu}_{B}^{d_1} \rightarrow \begin{cases}
        (0, 0)^T & x_0 \geq 0, y_0 \geq 0 \\
        (0, -2g_\alpha \alpha_y)^T & x_0 \geq 0, y_0 < 0 \\
        (-2g_\alpha \alpha_x, 0)^T & x_0 < 0, y_0 \geq 0 \\
        (-2g_\alpha \alpha_x, -2g_\alpha \alpha_y)^T & x_0 < 0, y_0 < 0. \\
    \end{cases}
\end{align}
As the strength $d$ of the classical modulation increases, the frequency of classical errors decreases and so the approximate transform $\hat D(-g_\alpha\langle \bm{\alpha} \rangle)$ rapidly approaches the correct displacement operation $\hat D(-g_\alpha \bm{ \alpha}_1 )$, such that
\begin{align}
    \bm{\mu}_{B}^{d_1} \longrightarrow (0,0)^T.
\end{align}

\subsection{Covariance matrix of the ensemble joint entangled state}
Since the bit-error rate of the protocol only achieves $e_C = 0$ in the limit of an infinitely strong classical signal, the output state $\hat \rho_{AB}^{d_1}$ of the protocol (after measurement and re-displacement) for a single shot is necessarily non-Gaussian. However, the degree of non-Gaussianity generally vanishes at least exponentially fast with the classical signal strength $d$. The output state is therefore well approximated by the equivalent Gaussian state with identical mean and covariance matrix.

In the simplified case where Alice sends only the single classical symbol $\tilde d_1$, the mean vector and covariance matrix of the ensemble state (i.e. in the limit of many shots, where each moment is averaged over the distribution of the measurement results $x_0, y_0$) is given by
\begin{align}
    \bm{\mu}_{AB}^{d_1} &= \left( 0, 0, 2 \sqrt{\tau} \alpha_x e_C, 2 \sqrt{\tau} \alpha_y e_C \right)^T \\
    V_{AB}^{d_1} &= \begin{pmatrix}
            a_{0} \id & c_0 \left[ 1 - \delta\right]\sz \\
            c_0 \left[ 1 - \delta \right]\sz & \left[b_0 + \Delta - \abs{\bm{\mu}_{B}^{d_1}}^2 \right] \id
        \end{pmatrix}.
\end{align}
We define
\begin{align}
    \delta &= \sqrt{\frac{2}{\pi}}\omega e^{-{\omega^2}/{2}} \\
    \Delta &= 2(1-b_{00}) \tau \delta + 2 \alpha^2\tau e_C
\end{align}
where $\omega$ is the signal-to-noise ratio of the Gaussian classical signal,
\begin{align}
    \omega^2 = \frac{1}{2}\frac{\alpha^2}{b_{00} + \cosh 2r_B}
\end{align}
such that
\begin{align}
    e_C &= \frac{1}{2} \mathrm{erfc} \left( \frac{\omega}{\sqrt{2}} \right).
\end{align}
Recall that $\alpha_x = \alpha_y = \alpha / \sqrt{2}$ for the symbol $\tilde d_1$. The variances $a_0, b_0$ and $c_0$ encode the variances of the quantum state in the case of no classical signal, i.e. the protocol where Alice and Bob perform only a partial CV teleportation with feed-forward on Bob's mode:
\begin{align}
    a_0 &= a_{00} \\
    b_0 &= \tau b_{00} + (1 - \tau) \\
    c_0 &= -\sqrt{\tau} c_{00}.
\end{align}
Extending to the full protocol, where Alice has essentially an equal chance of sending either of the four classical symbols, is straightforward. In this case, Alice's input state is not the single-symbol state $\hat \rho_{AA_S}^{d_1}$ but rather a balanced stochastic mix of each single-symbol state:
\begin{align}
    \hat \rho_{AA_S} &= \frac{\hat \rho_{AA_S}^{d_1} + \hat \rho_{AA_S}^{d_2} + \hat \rho_{AA_S}^{d_3} + \hat \rho_{AA_S}^{d_4}}{4}.
\end{align}
By the overall linearity of the protocol it follows that the output state is therefore also given by the equivalent mix of single-symbol output states:
\begin{align}
    \hat \rho_{AB}^\mathrm{out} &= \frac{\hat \rho_{AB}^{d_1} + \hat \rho_{AB}^{d_2} + \hat \rho_{AB}^{d_3} + \hat \rho_{AB}^{d_4}}{4}.
\end{align}
The general output state is also well approximated by a Gaussian for the same reasons as the single-symbol output state. Determining the covariance matrix and mean vector of the general output state is also straightforward, since we can exploit the linearity of expectation values; given the symmetry of the classical protocol we find
\begin{align}
    \bm{\mu}_{AB}^\mathrm{out} &= \left( 0, 0, 0, 0 \right)^T \\
    V_{AB}^\mathrm{out} &= \begin{pmatrix}
            a_{0} \id & c_0 \left[ 1 - \delta \right]\sz \\
            c_0 \left[ 1 - \delta \right]\sz & \left[b_0 + \Delta\right] \id
        \end{pmatrix}.
\end{align}
The two relevant limits \cite{zaunders_generalised_2026} are the limit of high teleporter squeezing $r_B \gg 0$ and high classical signal $\alpha \gg 0$. In the limit of large classical signal, it is straightforward to see that $\delta, \Delta \longrightarrow 0$, and so
\begin{align}
    V_{AB} &\longrightarrow \begin{pmatrix}
        a_0 \id & c_0 \sz \\
        c_0 \sz & b_0 \id
    \end{pmatrix}
\end{align}
i.e. perfect classical communication allows the two parties to recover a state equivalent to Alice's initial entangled state passed through the channel $\mathcal{E}$ and a pure-loss channel of transmissivity $\tau$. In the limit of high squeezing $r_B \longrightarrow \infty$ and fixed classical signal, however, we find $e_C \longrightarrow 0.5$ and
\begin{align}
    \delta &\longrightarrow 0 \\
    \Delta &\longrightarrow \alpha^2
\end{align}
as well as $b_{0}, c_0 \longrightarrow b_{00}, c_{00}$. Unlimited squeezing therefore allows Alice and Bob to ignore the effect of the teleporter at the cost of vanishing classical communication (due to the reduced signal-to-noise of Bob's measurements), the errors of which propagate an effective thermal-noise contribution $\alpha^2$ onto the quantum signal.

Finally, in the limit of both high classical signal and high squeezing, we obtain
\begin{align}
    V_{AB} &\longrightarrow \begin{pmatrix}
        a_{00} \id & c_{00} \sz \\
        c_{00} \sz & b_{00} \id
    \end{pmatrix}
\end{align}
i.e. they obtain an ensemble state exactly equivalent to if Alice had directly transmitted the entangled state $\hat \rho_A$ over the communication channel $\mathcal{E}$.

\section{Results} \label{sec:results}
We now proceed to characterise the protocol in terms of the entanglement of the non-Gaussian output state $\hat \rho_{AB}^\mathrm{out}$. We present results for three useful measures of mixed-state entanglement in continuous-variable systems: firstly, the reverse coherent information \cite{garcia-patron_reverse_2009}
\begin{align}
    \mathcal{R} = S(\hat \rho_A^\mathrm{out}) - S(\hat \rho_{AB}^\mathrm{out})
\end{align}
which provides a lower bound on the distillable entanglement of the output state \cite{weedbrook_gaussian_2012, winnel_generalized_2020, pirandola_advances_2020}; secondly, the entanglement of formation $\mathcal{E}_F$ \cite{bennett_mixed-state_1996, wolf_gaussian_2004}, which in a continuous-variable context quantifies the minimum amount of entanglement required to prepare $\hat \rho_{AB}^\mathrm{out}$ from vacuum \cite{wolf_extremality_2006, tserkis_quantifying_2017, tserkis_quantifying_2019}; and thirdly the asymptotic secret-key rate \cite{devetak_distillation_2005, laudenbach_continuousvariable_2018}
\begin{align}
    \mathcal{K}^\infty &= \beta I_{AB} - \chi_{EB}
\end{align}
which is a measure of the number of secret-key bits distillable by Alice and Bob from the state $\hat \rho_{AB}^\mathrm{out}$ in the limit of many states being exchanged.

\begin{figure*}
    \centering
    \includegraphics[width = \columnwidth]{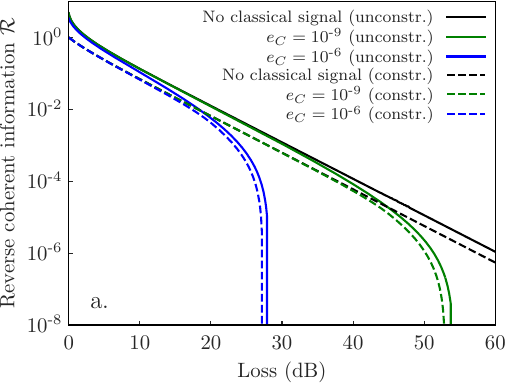}
    \hspace{10pt}
    \includegraphics[width = \columnwidth]{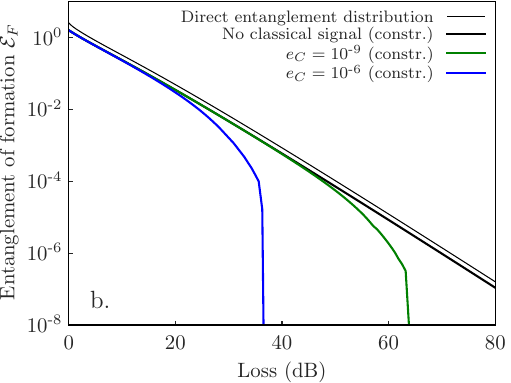}
    \caption{Lower bounds on the (a) reverse coherent information and (b) entanglement of formation of the generated state $\hat \rho_{AB}^\mathrm{out}$ generated by the protocol as a function of channel loss $\eta$. The excess noise on the channel is $\e = 0.025$. The reverse coherent information $\mathcal{R}$ is optimised over the squeezing of Alice and Bob's respective two-mode states $r_A, r_B$ at each point assuming unconstrained optimisation ($r_{A,B} \in [0,\infty)$, solid lines) and constrained optimisation ($r_{A,B} \in [0, 1.151]$, dashed lines) representing a maximum squeezing of 10 dB. The entanglement of formation $\mathcal{E}_F$ is optimised only for the constrained case, since the entanglement needed to prepare a state of unconstrained squeezing is trivially infinite \cite{tserkis_quantifying_2017}. Data is presented in both cases for a protocol with no classical signal (i.e. ordinary teleportation), $e_C = 10^{-9}$ and $e_C = 10^{-6}$.}
    \label{fig:pubfig_rci_eof}
\end{figure*}

For each measure $\mathcal{R}, \mathcal{E}_F, \mathcal{K}^\infty$, an exact analytical value for the full arbitrary non-Gaussian state $\hat \rho_{AB}^\mathrm{out}$ is generally not known. However, extremality properties of Gaussian states \cite{wolf_extremality_2006} allow us to derive a sufficiently tight lower bound on $\mathcal{R}, \mathcal{E}_F, \mathcal{K}^\infty$ by computing the value for the equivalent Gaussian state $\hat \rho_{AB}^G$ defined by the mean vector and covariance matrix $V_{AB}^\mathrm{out}$, $\bm \mu_{AB}^\mathrm{out}$. We present these calculations in Appendix \ref{app:ent}. Each bound is then optimised over Alice and Bob's two-mode squeezing parameters $r_{A,B}$ in both an impractical unconstrained regime, where it is assumed that Alice and Bob can generate states of essentially unbounded squeezing, and a constrained regime, where both parties are limited to a maximum squeezing of $10$ dB in order to capture a more realistic implementation of the protocol subject to limitations in state preparation \cite{vahlbruch_detection_2016}.

Additionally, we employ the approach outlined in refs. \cite{qi_noise_2018} and \cite{zaunders_enhanced_2025}, where we condition the performance of the protocol on a fixed classical quality-of-service metric. Specifically, we fix
\begin{align}
    d &= \frac{2 \sqrt{2} \sqrt{b_{00} + \cosh 2r_B} \ \mathrm{erfc}^{-1} \left( 2 e_C \right)}{\sqrt{\eta}}
\end{align}
such that the error rate of the classical data is always at most $e_C$. This more accurately reflects the usage of the protocol in actual deployment, where the classical aspect of the communications must obey strict requirements on reliability and performance, and where the energy cost associated with increasing the signal strength $d$ is not a chief concern.

Figures \ref{fig:pubfig_rci_eof}.a and \ref{fig:pubfig_rci_eof}.b show results for the reverse coherent information $\mathcal{R}$ and entanglement of formation $\mathcal{E}_F$ respectively in the case of no classical signal ($d = 0$, where the protocol reduces to forward-communication-assisted asymmetric entanglement swapping), $e_C = 10^{-9}$ and $e_C = 10^{-6}$. In the unconstrained regime for the protocol with no classical signal, the RCI approaches the repeaterless (PLOB) bound of $-\log_2(1 - \eta)$ \cite{pirandola_fundamental_2017}, which constitutes an ultimate upper bound on the entanglement distribution capacity of the protocol, since in the limit of unbounded squeezing the teleportation channel approaches identity. In the constrained regime the protocol achieves similar performance (i.e. scaling as $\mathcal{O}[\eta]$) in the limit of high loss, showing that Bob does not need to employ massive amounts of squeezing to achieve equivalent quantum performance.

The RCI for the protocol with a non-zero classical signal then demonstrates the decohering effect of the classical bit errors on the quantum signal, with the protocol only delivering distillable entanglement below $\sim 55$ dB of loss for a classical signal with $e_C = 10^{-9}$, and below $\sim 27$ dB of loss for a classical signal with $e_C = 10^{-6}$. The sensitivity of the Gaussian CMQC protocol to classical errors, in comparison to SQCC-QKD protocols \cite{zaunders_enhanced_2025} which are more or less robust to weak classical modulation, arises from the term $2 \alpha^2 \tau e_C$ in $\Delta$, which has has the equivalent effective excess-noise contribution $2d^2e_C$: even for fixed bit-error rates, as loss increases the classical amplitude $d$ required to maintain the signal increases exponentially and so $\Delta$ grows with loss. The protocol fails to deliver distillable entanglement when $\Delta$ becomes larger than the maximally tolerable excess noise (MTEN).

\begin{figure*}
    \centering
    \includegraphics[width = \columnwidth]{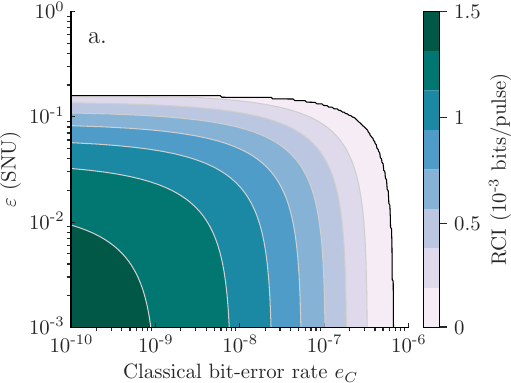}
    \hspace{10pt}
    \includegraphics[width = \columnwidth]{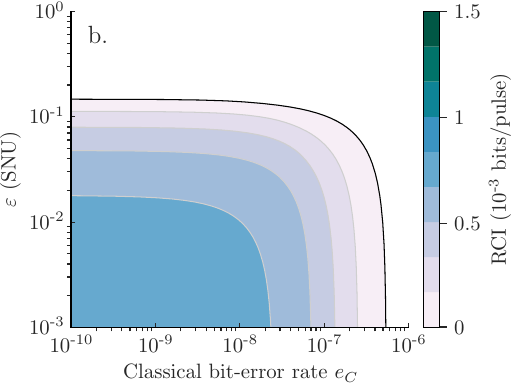}
    \caption{RCI $\mathcal{R}$ as a function of channel noise $\e$ and classical bit-error rate $e_C$, in the case of unconstrained (a) and constrained (b) optimisation for $\eta = 30$ dB channel loss. The boundary of the shaded area represents the maximum channel noise $\e_\mathrm{MTEN}$ at which the protocol can distil entanglement.}
    \label{fig:pubfig_rci_mten}
\end{figure*}

Figure \ref{fig:pubfig_rci_mten} characterises the protocol in terms of channel excess noise as a function of the classical quality-of-service for a fixed channel loss of $\eta = 30$ dB. We calculate the maximum tolerable excess noise $\e_\mathrm{MTEN}$ in terms of the RCI for both the unconstrained and constrained regimes. In the unconstrained regime (Fig. \ref{fig:pubfig_rci_mten}.a), the sensitivity to classical noise is highlighted more clearly, with the classical signal requiring a bit-error rate of at most $e_C \sim 5\times10^{-7}$ to be able to distil entanglement. However, the transition between the distillable and non-distillable regime is sharp, which shows there is very little coupling between channel noise and classical noise. In the distillable regime, the maximum allowable noise stays mostly constant at $\e \sim 10^{-1}$, which indicates the protocol is fairly robust over noisy channels.

We also see again that the difference between the unconstrained and constrained (Fig. \ref{fig:pubfig_rci_mten}.a, b respectively) protocols is small, with the constrained protocol exhibiting the same $\e_\mathrm{MTEN}$ as a function of $e_C$ as in the unconstrained case. The only disadvantage incurred in the constrained case is a smaller magnitude of entanglement for lower bit-error rates.

\begin{figure}
    \centering
    \includegraphics[width = \columnwidth]{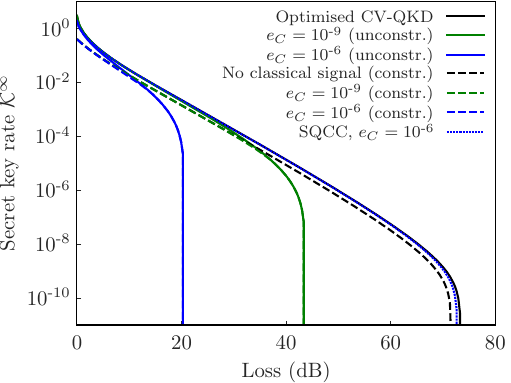}
    \caption{Secret-key rate $\mathcal{K}^\infty$ as a function of channel loss for the same parameters as in Fig. \ref{fig:pubfig_rci_eof}. The key rate of the protocol with no classical signal and unconstrained optimisation is equivalent to the optimised key rate for ordinary Gaussian coherent-state CV-QKD performed over $\mathcal{E}$, which upper-bounds the performance of the protocol. For comparison, we include a plot of the keyrate attained by the improved SQCC-QKD protocol described in \cite{zaunders_enhanced_2025}, which nearly saturates the bound despite the relatively poor ($e_C = 10^{-6}$) classical signal.}
    \label{fig:pubfig_qkd_asym}
\end{figure}

Lastly, we present the asymptotic secret key generation rate results in Figure \ref{fig:pubfig_qkd_asym} for unconstrained and constrained optimisation, again for no classical signal, $e_C = 10^{-9}$, and $e_C = 10^{-6}$. The rate $\mathcal{K}^\infty$ follows the same trends as in the distillable entanglement, and exhibits the same characteristic sensitivity to classical noise. Importantly, Fig. \ref{fig:pubfig_qkd_asym} also distinguishes between the entanglement-based protocol and the SQCC scheme presented in \cite{zaunders_enhanced_2025}, which uses classical modulation of coherent states and postprocessed direct measurement to generate secret-key bits at the same time as exchanging classical information. Compared to the QKD-specific SQCC scheme, the entanglement scheme is significantly more sensitive to errors in the classical signal: the upper bound on key generation, which is given by the optimised secret-key rate of the channel in the case of no classical signal, is saturated by the optimised SQCC-QKD protocol at bit-error rates of only $e_C = 10^{-6}$. In contrast, the Gaussian CMQC scheme is less robust, requiring error rates of up to $e_C \leq 10^{-14}$ to saturate the upper bound. The reason for this is that the re-displacement process in SQCC-QKD is done virtually on measurement outcomes, and so is not constrained to be a physically valid operation, unlike the re-displacement $\hat D(-g_\alpha\langle\bm{\alpha}\rangle)$. The virtual re-displacement therefore has the effect of reducing the effective noise on Bob's mode, something which is not possible with physical operations, and so after normalisation does not have an effective SNR-dependent noise contribution. This allows the SQCC-QKD protocol to be much more resilient to classical errors compared to the physical CMQC protocol.

However, it is worth noting that the approach to QKD used here is pessimistic, since it implicitly assumes Eve has knowledge of the measurement result $\bm x$, i.e. that the teleportation step is untrusted. It is also interesting to note that the Gaussian CMQC protocol, which can be represented in terms of some physical (albeit non-Gaussian) effective channel \cite{zaunders_generalised_2026}, offers an alternative proof of the security of CV-QKD and simultaneous classical communication when used in this way. In ref. \cite{zaunders_enhanced_2025}, the non-physicality of the re-displacement operations required the use of some cumbersome security arguments to lower-bound the information of an attacker on the quantum key data. The key rate of the CMQC protocol, while not as tight, requires fewer assumptions to guarantee security and likely could be improved further by relaxing requirements on what aspects of the teleportation are assumed to be controlled by Eve.

\section{Discussion} \label{sec:discussion}
Here we have discussed how the CMQC protocol of ref. \cite{zaunders_generalised_2026} can be implemented to efficiently distribute both classical information and continuous-variable Gaussian entanglement on the same pulse and optical mode. The protocol leverages a CV teleportation of the hybrid quantum-classical signal to both measure the classical data and preserve the quantum correlations, and additionally uses the measurement outcomes of the teleportation to approximate a pure Gaussian entangled state via a local unitary operation. We provide an exact analytic description of the entangled state shared by Alice and Bob after performing the protocol over a lossy, noisy Gaussian channel and quantify the bit-error rate of the classical signal. Lastly, we assess the efficacy of the entanglement distribution by computing the entanglement of the output state in terms of the reverse coherent information, entanglement of formation and asymptotic secret-key generation rate; we compare the performance of the protocol to the point-to-point SQCC CV-QKD scheme developed by Qi et al. \cite{qi_simultaneous_2016, zaunders_enhanced_2025}, which it is virtually equivalent to. We show that the entanglement distribution protocol, while sensitive to classical measurement errors, is nonetheless robust against loss and channel noise, and retains this robustness even in the case of implementation under realistic conditions (i.e. imperfect teleportation). 

The application of the Gaussian CMQC scheme to quantum networking is obvious. The most immediately apparent use case of the protocol is the augmentation of existing or future optical-frequency classical communications networks, where a network of many classically-communicating nodes now promotes each outgoing classical signal to a hybrid quantum-classical signal via the Gaussian CMQC protocol. Each node retains the ability to send and receive classical information without requiring any changes to the encoding, decoding or synchronisation of each coherent signal, only requiring the addition of an entanglement source (for transmission) and teleportation apparatus (for receiving). Combined with an appropriate distillation protocol \cite{guanzon_ideal_2022, dias_quantum_2020, azuma_quantum_2023} and quantum memories, this would allow each node in the system to build up a reservoir of entangled pairs with every other networked node, in the background, while maintaining uptime on each classical link. Each node can then leverage this stored entanglement to perform whatever task is required, e.g. entanglement swapping between distant nodes, key distribution, distributed quantum computing, etc., and can perform the additional classical communications needed for each task seamlessly via the same channel. 

It would also be straightforward to extend the Gaussian CMQC protocol to either a two-way or repeater-like scheme. In the case of a two-way scheme, individual signals could be exchanged in both the forward and backward direction between Alice and Bob, allowing for parties to engage in two-way classical communication while simultaneously generating entanglement. This would have the advantage of also allowing Alice and Bob to mutually perform feedback operations on their retained entangled mode, which allows them to (in principle) achieve a pure state under appropriate conditions \cite{braunstein_quantum_2005}, as well as letting them implement higher-rate distillation protocols which require two-way or backwards classical communication. Similarly, because the protocol outputs some non-pure zero-mean CV entangled state, it is possible to extend the Gaussian CMQC scheme to a repeater-like configuration by purifying the state through noiseless linear amplification \cite{ralph_nondeterministic_2009, xiang_heralded_2010}, which would have the effect of increasing the effective squeezing of the joint entangled state. The classical signal, which is measured independently, can also be error-corrected and the outgoing mode then re-displaced with the appropriate classical symbol. In this way, the hybrid quantum-classical signal could be propagated indefinitely in one direction, subject only to limitations arising from the probabilistic nature of quantum amplifiers. Alternatively, the scheme could be symmetrised to output a hybrid signal to two nodes simultaneously, allowing individual CMQC `units' to form part of a combined quantum-classical repeater protocol of the kind described by Dias et al. \cite{dias_distributing_2022}, though this may also be subject to issues of synchronisation (see e.g. \cite{rohde_quantum_2025}) of both the classical and quantum signal.

Lastly, the teleportation protocol \cite{braunstein_quantum_2005} used in this paper is the simplest possible CV teleportation scheme, and is intended primarily as a proof-of-concept. Other proposals for high-fidelity CV teleportation exist, e.g. \cite{andersen_high-fidelity_2013}, and it would be valuable to determine whether implementing these more sophisticated teleportation schemes within the overall Gaussian CMQC protocol may lead to improved outcomes either in classical quality-of-service or in terms of the generated entangled state.

\begin{acknowledgments}
We acknowledge productive discussions with Josephine Dias and Nibedita Swain. The Australian Government supported this research through the Australian Research Council’s Linkage Projects funding scheme (Project No. LP200100601) and the Centre of Excellence for Quantum Computation and Communication Technology (Project No. CE170100012). The views expressed herein are those of the authors and are not necessarily those of the Australian Government or the Australian Research Council.
\end{acknowledgments}


\begin{thebibliography}{62}%
\makeatletter
\providecommand \@ifxundefined [1]{%
 \@ifx{#1\undefined}
}%
\providecommand \@ifnum [1]{%
 \ifnum #1\expandafter \@firstoftwo
 \else \expandafter \@secondoftwo
 \fi
}%
\providecommand \@ifx [1]{%
 \ifx #1\expandafter \@firstoftwo
 \else \expandafter \@secondoftwo
 \fi
}%
\providecommand \natexlab [1]{#1}%
\providecommand \enquote  [1]{``#1''}%
\providecommand \bibnamefont  [1]{#1}%
\providecommand \bibfnamefont [1]{#1}%
\providecommand \citenamefont [1]{#1}%
\providecommand \href@noop [0]{\@secondoftwo}%
\providecommand \href [0]{\begingroup \@sanitize@url \@href}%
\providecommand \@href[1]{\@@startlink{#1}\@@href}%
\providecommand \@@href[1]{\endgroup#1\@@endlink}%
\providecommand \@sanitize@url [0]{\catcode `\\12\catcode `\$12\catcode `\&12\catcode `\#12\catcode `\^12\catcode `\_12\catcode `\%12\relax}%
\providecommand \@@startlink[1]{}%
\providecommand \@@endlink[0]{}%
\providecommand \url  [0]{\begingroup\@sanitize@url \@url }%
\providecommand \@url [1]{\endgroup\@href {#1}{\urlprefix }}%
\providecommand \urlprefix  [0]{URL }%
\providecommand \Eprint [0]{\href }%
\providecommand \doibase [0]{https://doi.org/}%
\providecommand \selectlanguage [0]{\@gobble}%
\providecommand \bibinfo  [0]{\@secondoftwo}%
\providecommand \bibfield  [0]{\@secondoftwo}%
\providecommand \translation [1]{[#1]}%
\providecommand \BibitemOpen [0]{}%
\providecommand \bibitemStop [0]{}%
\providecommand \bibitemNoStop [0]{.\EOS\space}%
\providecommand \EOS [0]{\spacefactor3000\relax}%
\providecommand \BibitemShut  [1]{\csname bibitem#1\endcsname}%
\let\auto@bib@innerbib\@empty
\bibitem [{\citenamefont {Zaunders}\ and\ \citenamefont {Ralph}(2026)}]{zaunders_generalised_2026}%
  \BibitemOpen
  \bibfield  {author} {\bibinfo {author} {\bibfnamefont {N.}~\bibnamefont {Zaunders}}\ and\ \bibinfo {author} {\bibfnamefont {T.~C.}\ \bibnamefont {Ralph}},\ }\href {https://doi.org/10.48550/arXiv.2606.03181} {\bibinfo {title} {Generalised simultaneous transmission of arbitrary quantum states and classical information}} (\bibinfo {year} {2026}),\ \bibinfo {note} {arXiv:2606.03181 [quant-ph]}\BibitemShut {NoStop}%
\bibitem [{\citenamefont {Nielsen}\ and\ \citenamefont {Chuang}(2010)}]{nielsen_quantum_2010}%
  \BibitemOpen
  \bibfield  {author} {\bibinfo {author} {\bibfnamefont {M.~A.}\ \bibnamefont {Nielsen}}\ and\ \bibinfo {author} {\bibfnamefont {I.~L.}\ \bibnamefont {Chuang}},\ }\href@noop {} {{\selectlanguage {en}\emph {\bibinfo {title} {Quantum computation and quantum information}}}},\ \bibinfo {edition} {10th}\ ed.\ (\bibinfo  {publisher} {Cambridge University Press},\ \bibinfo {address} {Cambridge ; New York},\ \bibinfo {year} {2010})\BibitemShut {NoStop}%
\bibitem [{\citenamefont {Bennett}\ and\ \citenamefont {Brassard}(2014)}]{bennett_quantum_2014}%
  \BibitemOpen
  \bibfield  {author} {\bibinfo {author} {\bibfnamefont {C.~H.}\ \bibnamefont {Bennett}}\ and\ \bibinfo {author} {\bibfnamefont {G.}~\bibnamefont {Brassard}},\ }\bibfield  {title} {{\selectlanguage {en}\bibinfo {title} {Quantum cryptography: {Public} key distribution and coin tossing}},\ }\href {https://doi.org/10.1016/j.tcs.2014.05.025} {\bibfield  {journal} {\bibinfo  {journal} {Theoretical Computer Science}\ }\textbf {\bibinfo {volume} {560}},\ \bibinfo {pages} {7} (\bibinfo {year} {2014})}\BibitemShut {NoStop}%
\bibitem [{\citenamefont {Ekert}(1991)}]{ekert_quantum_1991}%
  \BibitemOpen
  \bibfield  {author} {\bibinfo {author} {\bibfnamefont {A.~K.}\ \bibnamefont {Ekert}},\ }\bibfield  {title} {{\selectlanguage {en}\bibinfo {title} {Quantum cryptography based on {Bell}’s theorem}},\ }\href {https://doi.org/10.1103/PhysRevLett.67.661} {\bibfield  {journal} {\bibinfo  {journal} {Physical Review Letters}\ }\textbf {\bibinfo {volume} {67}},\ \bibinfo {pages} {661} (\bibinfo {year} {1991})}\BibitemShut {NoStop}%
\bibitem [{\citenamefont {Gisin}\ \emph {et~al.}(2002)\citenamefont {Gisin}, \citenamefont {Ribordy}, \citenamefont {Tittel},\ and\ \citenamefont {Zbinden}}]{gisin_quantum_2002}%
  \BibitemOpen
  \bibfield  {author} {\bibinfo {author} {\bibfnamefont {N.}~\bibnamefont {Gisin}}, \bibinfo {author} {\bibfnamefont {G.}~\bibnamefont {Ribordy}}, \bibinfo {author} {\bibfnamefont {W.}~\bibnamefont {Tittel}},\ and\ \bibinfo {author} {\bibfnamefont {H.}~\bibnamefont {Zbinden}},\ }\bibfield  {title} {{\selectlanguage {en}\bibinfo {title} {Quantum cryptography}},\ }\href {https://doi.org/10.1103/RevModPhys.74.145} {\bibfield  {journal} {\bibinfo  {journal} {Reviews of Modern Physics}\ }\textbf {\bibinfo {volume} {74}},\ \bibinfo {pages} {145} (\bibinfo {year} {2002})}\BibitemShut {NoStop}%
\bibitem [{\citenamefont {Ralph}(1999)}]{ralph_continuous_1999}%
  \BibitemOpen
  \bibfield  {author} {\bibinfo {author} {\bibfnamefont {T.~C.}\ \bibnamefont {Ralph}},\ }\bibfield  {title} {{\selectlanguage {en}\bibinfo {title} {Continuous variable quantum cryptography}},\ }\href {https://doi.org/10.1103/PhysRevA.61.010303} {\bibfield  {journal} {\bibinfo  {journal} {Physical Review A}\ }\textbf {\bibinfo {volume} {61}},\ \bibinfo {pages} {010303} (\bibinfo {year} {1999})}\BibitemShut {NoStop}%
\bibitem [{\citenamefont {Grosshans}\ \emph {et~al.}(2003)\citenamefont {Grosshans}, \citenamefont {Van~Assche}, \citenamefont {Wenger}, \citenamefont {Brouri}, \citenamefont {Cerf},\ and\ \citenamefont {Grangier}}]{grosshans_quantum_2003}%
  \BibitemOpen
  \bibfield  {author} {\bibinfo {author} {\bibfnamefont {F.}~\bibnamefont {Grosshans}}, \bibinfo {author} {\bibfnamefont {G.}~\bibnamefont {Van~Assche}}, \bibinfo {author} {\bibfnamefont {J.}~\bibnamefont {Wenger}}, \bibinfo {author} {\bibfnamefont {R.}~\bibnamefont {Brouri}}, \bibinfo {author} {\bibfnamefont {N.~J.}\ \bibnamefont {Cerf}},\ and\ \bibinfo {author} {\bibfnamefont {P.}~\bibnamefont {Grangier}},\ }\bibfield  {title} {{\selectlanguage {en}\bibinfo {title} {Quantum key distribution using gaussian-modulated coherent states}},\ }\href {https://doi.org/10.1038/nature01289} {\bibfield  {journal} {\bibinfo  {journal} {Nature}\ }\textbf {\bibinfo {volume} {421}},\ \bibinfo {pages} {238} (\bibinfo {year} {2003})}\BibitemShut {NoStop}%
\bibitem [{\citenamefont {Bennett}\ \emph {et~al.}(1993)\citenamefont {Bennett}, \citenamefont {Brassard}, \citenamefont {Crépeau}, \citenamefont {Jozsa}, \citenamefont {Peres},\ and\ \citenamefont {Wootters}}]{bennett_teleporting_1993}%
  \BibitemOpen
  \bibfield  {author} {\bibinfo {author} {\bibfnamefont {C.~H.}\ \bibnamefont {Bennett}}, \bibinfo {author} {\bibfnamefont {G.}~\bibnamefont {Brassard}}, \bibinfo {author} {\bibfnamefont {C.}~\bibnamefont {Crépeau}}, \bibinfo {author} {\bibfnamefont {R.}~\bibnamefont {Jozsa}}, \bibinfo {author} {\bibfnamefont {A.}~\bibnamefont {Peres}},\ and\ \bibinfo {author} {\bibfnamefont {W.~K.}\ \bibnamefont {Wootters}},\ }\bibfield  {title} {{\selectlanguage {en}\bibinfo {title} {Teleporting an unknown quantum state via dual classical and {Einstein}-{Podolsky}-{Rosen} channels}},\ }\href {https://doi.org/10.1103/PhysRevLett.70.1895} {\bibfield  {journal} {\bibinfo  {journal} {Physical Review Letters}\ }\textbf {\bibinfo {volume} {70}},\ \bibinfo {pages} {1895} (\bibinfo {year} {1993})}\BibitemShut {NoStop}%
\bibitem [{\citenamefont {Bennett}\ \emph {et~al.}(1996{\natexlab{a}})\citenamefont {Bennett}, \citenamefont {Brassard}, \citenamefont {Popescu}, \citenamefont {Schumacher}, \citenamefont {Smolin},\ and\ \citenamefont {Wootters}}]{bennett_purification_1996}%
  \BibitemOpen
  \bibfield  {author} {\bibinfo {author} {\bibfnamefont {C.~H.}\ \bibnamefont {Bennett}}, \bibinfo {author} {\bibfnamefont {G.}~\bibnamefont {Brassard}}, \bibinfo {author} {\bibfnamefont {S.}~\bibnamefont {Popescu}}, \bibinfo {author} {\bibfnamefont {B.}~\bibnamefont {Schumacher}}, \bibinfo {author} {\bibfnamefont {J.~A.}\ \bibnamefont {Smolin}},\ and\ \bibinfo {author} {\bibfnamefont {W.~K.}\ \bibnamefont {Wootters}},\ }\bibfield  {title} {{\selectlanguage {en}\bibinfo {title} {Purification of {Noisy} {Entanglement} and {Faithful} {Teleportation} via {Noisy} {Channels}}},\ }\href {https://doi.org/10.1103/PhysRevLett.76.722} {\bibfield  {journal} {\bibinfo  {journal} {Physical Review Letters}\ }\textbf {\bibinfo {volume} {76}},\ \bibinfo {pages} {722} (\bibinfo {year} {1996}{\natexlab{a}})}\BibitemShut {NoStop}%
\bibitem [{\citenamefont {Braunstein}\ \emph {et~al.}(2000)\citenamefont {Braunstein}, \citenamefont {Fuchs},\ and\ \citenamefont {Kimble}}]{braunstein_criteria_2000}%
  \BibitemOpen
  \bibfield  {author} {\bibinfo {author} {\bibfnamefont {S.~L.}\ \bibnamefont {Braunstein}}, \bibinfo {author} {\bibfnamefont {C.~A.}\ \bibnamefont {Fuchs}},\ and\ \bibinfo {author} {\bibfnamefont {H.~J.}\ \bibnamefont {Kimble}},\ }\bibfield  {title} {{\selectlanguage {en}\bibinfo {title} {Criteria for continuous-variable quantum teleportation}},\ }\href {https://doi.org/10.1080/09500340008244041} {\bibfield  {journal} {\bibinfo  {journal} {Journal of Modern Optics}\ }\textbf {\bibinfo {volume} {47}},\ \bibinfo {pages} {267} (\bibinfo {year} {2000})}\BibitemShut {NoStop}%
\bibitem [{\citenamefont {Jozsa}\ \emph {et~al.}(2000)\citenamefont {Jozsa}, \citenamefont {Abrams}, \citenamefont {Dowling},\ and\ \citenamefont {Williams}}]{jozsa_quantum_2000}%
  \BibitemOpen
  \bibfield  {author} {\bibinfo {author} {\bibfnamefont {R.}~\bibnamefont {Jozsa}}, \bibinfo {author} {\bibfnamefont {D.~S.}\ \bibnamefont {Abrams}}, \bibinfo {author} {\bibfnamefont {J.~P.}\ \bibnamefont {Dowling}},\ and\ \bibinfo {author} {\bibfnamefont {C.~P.}\ \bibnamefont {Williams}},\ }\bibfield  {title} {{\selectlanguage {en}\bibinfo {title} {Quantum {Clock} {Synchronization} {Based} on {Shared} {Prior} {Entanglement}}},\ }\href {https://doi.org/10.1103/PhysRevLett.85.2010} {\bibfield  {journal} {\bibinfo  {journal} {Physical Review Letters}\ }\textbf {\bibinfo {volume} {85}},\ \bibinfo {pages} {2010} (\bibinfo {year} {2000})}\BibitemShut {NoStop}%
\bibitem [{\citenamefont {Yurtsever}\ and\ \citenamefont {Dowling}(2002)}]{yurtsever_lorentz-invariant_2002}%
  \BibitemOpen
  \bibfield  {author} {\bibinfo {author} {\bibfnamefont {U.}~\bibnamefont {Yurtsever}}\ and\ \bibinfo {author} {\bibfnamefont {J.~P.}\ \bibnamefont {Dowling}},\ }\bibfield  {title} {{\selectlanguage {en}\bibinfo {title} {Lorentz-invariant look at quantum clock-synchronization protocols based on distributed entanglement}},\ }\href {https://doi.org/10.1103/PhysRevA.65.052317} {\bibfield  {journal} {\bibinfo  {journal} {Physical Review A}\ }\textbf {\bibinfo {volume} {65}},\ \bibinfo {pages} {052317} (\bibinfo {year} {2002})}\BibitemShut {NoStop}%
\bibitem [{\citenamefont {Kómár}\ \emph {et~al.}(2014)\citenamefont {Kómár}, \citenamefont {Kessler}, \citenamefont {Bishof}, \citenamefont {Jiang}, \citenamefont {Sørensen}, \citenamefont {Ye},\ and\ \citenamefont {Lukin}}]{komar_quantum_2014}%
  \BibitemOpen
  \bibfield  {author} {\bibinfo {author} {\bibfnamefont {P.}~\bibnamefont {Kómár}}, \bibinfo {author} {\bibfnamefont {E.~M.}\ \bibnamefont {Kessler}}, \bibinfo {author} {\bibfnamefont {M.}~\bibnamefont {Bishof}}, \bibinfo {author} {\bibfnamefont {L.}~\bibnamefont {Jiang}}, \bibinfo {author} {\bibfnamefont {A.~S.}\ \bibnamefont {Sørensen}}, \bibinfo {author} {\bibfnamefont {J.}~\bibnamefont {Ye}},\ and\ \bibinfo {author} {\bibfnamefont {M.~D.}\ \bibnamefont {Lukin}},\ }\bibfield  {title} {{\selectlanguage {en}\bibinfo {title} {A quantum network of clocks}},\ }\href {https://doi.org/10.1038/nphys3000} {\bibfield  {journal} {\bibinfo  {journal} {Nature Physics}\ }\textbf {\bibinfo {volume} {10}},\ \bibinfo {pages} {582} (\bibinfo {year} {2014})}\BibitemShut {NoStop}%
\bibitem [{\citenamefont {Ilo-Okeke}\ \emph {et~al.}(2018)\citenamefont {Ilo-Okeke}, \citenamefont {Tessler}, \citenamefont {Dowling},\ and\ \citenamefont {Byrnes}}]{ilo-okeke_remote_2018}%
  \BibitemOpen
  \bibfield  {author} {\bibinfo {author} {\bibfnamefont {E.~O.}\ \bibnamefont {Ilo-Okeke}}, \bibinfo {author} {\bibfnamefont {L.}~\bibnamefont {Tessler}}, \bibinfo {author} {\bibfnamefont {J.~P.}\ \bibnamefont {Dowling}},\ and\ \bibinfo {author} {\bibfnamefont {T.}~\bibnamefont {Byrnes}},\ }\bibfield  {title} {{\selectlanguage {en}\bibinfo {title} {Remote quantum clock synchronization without synchronized clocks}},\ }\href {https://doi.org/10.1038/s41534-018-0090-2} {\bibfield  {journal} {\bibinfo  {journal} {npj Quantum Information}\ }\textbf {\bibinfo {volume} {4}},\ \bibinfo {pages} {40} (\bibinfo {year} {2018})}\BibitemShut {NoStop}%
\bibitem [{\citenamefont {Cirac}\ \emph {et~al.}(1999)\citenamefont {Cirac}, \citenamefont {Ekert}, \citenamefont {Huelga},\ and\ \citenamefont {Macchiavello}}]{cirac_distributed_1999}%
  \BibitemOpen
  \bibfield  {author} {\bibinfo {author} {\bibfnamefont {J.~I.}\ \bibnamefont {Cirac}}, \bibinfo {author} {\bibfnamefont {A.~K.}\ \bibnamefont {Ekert}}, \bibinfo {author} {\bibfnamefont {S.~F.}\ \bibnamefont {Huelga}},\ and\ \bibinfo {author} {\bibfnamefont {C.}~\bibnamefont {Macchiavello}},\ }\bibfield  {title} {{\selectlanguage {en}\bibinfo {title} {Distributed quantum computation over noisy channels}},\ }\href {https://doi.org/10.1103/PhysRevA.59.4249} {\bibfield  {journal} {\bibinfo  {journal} {Physical Review A}\ }\textbf {\bibinfo {volume} {59}},\ \bibinfo {pages} {4249} (\bibinfo {year} {1999})}\BibitemShut {NoStop}%
\bibitem [{\citenamefont {Gottesman}\ \emph {et~al.}(2012)\citenamefont {Gottesman}, \citenamefont {Jennewein},\ and\ \citenamefont {Croke}}]{gottesman_longer-baseline_2012}%
  \BibitemOpen
  \bibfield  {author} {\bibinfo {author} {\bibfnamefont {D.}~\bibnamefont {Gottesman}}, \bibinfo {author} {\bibfnamefont {T.}~\bibnamefont {Jennewein}},\ and\ \bibinfo {author} {\bibfnamefont {S.}~\bibnamefont {Croke}},\ }\bibfield  {title} {{\selectlanguage {en}\bibinfo {title} {Longer-{Baseline} {Telescopes} {Using} {Quantum} {Repeaters}}},\ }\href {https://doi.org/10.1103/PhysRevLett.109.070503} {\bibfield  {journal} {\bibinfo  {journal} {Physical Review Letters}\ }\textbf {\bibinfo {volume} {109}},\ \bibinfo {pages} {070503} (\bibinfo {year} {2012})}\BibitemShut {NoStop}%
\bibitem [{\citenamefont {Degen}\ \emph {et~al.}(2017)\citenamefont {Degen}, \citenamefont {Reinhard},\ and\ \citenamefont {Cappellaro}}]{degen_quantum_2017}%
  \BibitemOpen
  \bibfield  {author} {\bibinfo {author} {\bibfnamefont {C.}~\bibnamefont {Degen}}, \bibinfo {author} {\bibfnamefont {F.}~\bibnamefont {Reinhard}},\ and\ \bibinfo {author} {\bibfnamefont {P.}~\bibnamefont {Cappellaro}},\ }\bibfield  {title} {{\selectlanguage {en}\bibinfo {title} {Quantum sensing}},\ }\href {https://doi.org/10.1103/RevModPhys.89.035002} {\bibfield  {journal} {\bibinfo  {journal} {Reviews of Modern Physics}\ }\textbf {\bibinfo {volume} {89}},\ \bibinfo {pages} {035002} (\bibinfo {year} {2017})}\BibitemShut {NoStop}%
\bibitem [{\citenamefont {Zhuang}\ \emph {et~al.}(2018)\citenamefont {Zhuang}, \citenamefont {Zhang},\ and\ \citenamefont {Shapiro}}]{zhuang_distributed_2018}%
  \BibitemOpen
  \bibfield  {author} {\bibinfo {author} {\bibfnamefont {Q.}~\bibnamefont {Zhuang}}, \bibinfo {author} {\bibfnamefont {Z.}~\bibnamefont {Zhang}},\ and\ \bibinfo {author} {\bibfnamefont {J.~H.}\ \bibnamefont {Shapiro}},\ }\bibfield  {title} {{\selectlanguage {en}\bibinfo {title} {Distributed quantum sensing using continuous-variable multipartite entanglement}},\ }\href {https://doi.org/10.1103/PhysRevA.97.032329} {\bibfield  {journal} {\bibinfo  {journal} {Physical Review A}\ }\textbf {\bibinfo {volume} {97}},\ \bibinfo {pages} {032329} (\bibinfo {year} {2018})}\BibitemShut {NoStop}%
\bibitem [{\citenamefont {Xia}\ \emph {et~al.}(2019)\citenamefont {Xia}, \citenamefont {Zhuang}, \citenamefont {Clark},\ and\ \citenamefont {Zhang}}]{xia_repeater-enhanced_2019}%
  \BibitemOpen
  \bibfield  {author} {\bibinfo {author} {\bibfnamefont {Y.}~\bibnamefont {Xia}}, \bibinfo {author} {\bibfnamefont {Q.}~\bibnamefont {Zhuang}}, \bibinfo {author} {\bibfnamefont {W.}~\bibnamefont {Clark}},\ and\ \bibinfo {author} {\bibfnamefont {Z.}~\bibnamefont {Zhang}},\ }\bibfield  {title} {{\selectlanguage {en}\bibinfo {title} {Repeater-enhanced distributed quantum sensing based on continuous-variable multipartite entanglement}},\ }\href {https://doi.org/10.1103/PhysRevA.99.012328} {\bibfield  {journal} {\bibinfo  {journal} {Physical Review A}\ }\textbf {\bibinfo {volume} {99}},\ \bibinfo {pages} {012328} (\bibinfo {year} {2019})}\BibitemShut {NoStop}%
\bibitem [{\citenamefont {Walls}\ and\ \citenamefont {Milburn}(2008)}]{walls_quantum_2008}%
  \BibitemOpen
  \bibfield  {author} {\bibinfo {author} {\bibfnamefont {D.~F.}\ \bibnamefont {Walls}}\ and\ \bibinfo {author} {\bibfnamefont {G.~J.}\ \bibnamefont {Milburn}},\ }\href@noop {} {{\selectlanguage {en}\emph {\bibinfo {title} {Quantum optics}}}},\ \bibinfo {edition} {2nd}\ ed.\ (\bibinfo  {publisher} {Springer},\ \bibinfo {address} {Berlin},\ \bibinfo {year} {2008})\BibitemShut {NoStop}%
\bibitem [{\citenamefont {Gerry}\ and\ \citenamefont {Knight}(2004)}]{gerry_introductory_2004}%
  \BibitemOpen
  \bibfield  {author} {\bibinfo {author} {\bibfnamefont {C.}~\bibnamefont {Gerry}}\ and\ \bibinfo {author} {\bibfnamefont {P.}~\bibnamefont {Knight}},\ }\href {https://doi.org/10.1017/CBO9780511791239} {\emph {\bibinfo {title} {Introductory {Quantum} {Optics}}}},\ \bibinfo {edition} {1st}\ ed.\ (\bibinfo  {publisher} {Cambridge University Press},\ \bibinfo {year} {2004})\BibitemShut {NoStop}%
\bibitem [{\citenamefont {Bachor}\ and\ \citenamefont {Ralph}(2019)}]{bachor_guide_2019}%
  \BibitemOpen
  \bibfield  {author} {\bibinfo {author} {\bibfnamefont {H.}~\bibnamefont {Bachor}}\ and\ \bibinfo {author} {\bibfnamefont {T.~C.}\ \bibnamefont {Ralph}},\ }\href {https://doi.org/10.1002/9783527695805} {{\selectlanguage {en}\emph {\bibinfo {title} {A {Guide} to {Experiments} in {Quantum} {Optics}}}}},\ \bibinfo {edition} {1st}\ ed.\ (\bibinfo  {publisher} {Wiley},\ \bibinfo {year} {2019})\BibitemShut {NoStop}%
\bibitem [{\citenamefont {Usenko}\ \emph {et~al.}(2025)\citenamefont {Usenko}, \citenamefont {Acín}, \citenamefont {Alléaume}, \citenamefont {Andersen}, \citenamefont {Diamanti}, \citenamefont {Gehring}, \citenamefont {Hajomer}, \citenamefont {Kanitschar}, \citenamefont {Pacher}, \citenamefont {Pirandola},\ and\ \citenamefont {Pruneri}}]{usenko_continuous-variable_2025}%
  \BibitemOpen
  \bibfield  {author} {\bibinfo {author} {\bibfnamefont {V.~C.}\ \bibnamefont {Usenko}}, \bibinfo {author} {\bibfnamefont {A.}~\bibnamefont {Acín}}, \bibinfo {author} {\bibfnamefont {R.}~\bibnamefont {Alléaume}}, \bibinfo {author} {\bibfnamefont {U.~L.}\ \bibnamefont {Andersen}}, \bibinfo {author} {\bibfnamefont {E.}~\bibnamefont {Diamanti}}, \bibinfo {author} {\bibfnamefont {T.}~\bibnamefont {Gehring}}, \bibinfo {author} {\bibfnamefont {A.~A.~E.}\ \bibnamefont {Hajomer}}, \bibinfo {author} {\bibfnamefont {F.}~\bibnamefont {Kanitschar}}, \bibinfo {author} {\bibfnamefont {C.}~\bibnamefont {Pacher}}, \bibinfo {author} {\bibfnamefont {S.}~\bibnamefont {Pirandola}},\ and\ \bibinfo {author} {\bibfnamefont {V.}~\bibnamefont {Pruneri}},\ }\href {https://doi.org/10.48550/ARXIV.2501.12801} {\bibinfo {title} {Continuous-variable quantum communication}} (\bibinfo {year} {2025}),\ \bibinfo {note} {version Number: 1}\BibitemShut {NoStop}%
\bibitem [{\citenamefont {Qi}(2016)}]{qi_simultaneous_2016}%
  \BibitemOpen
  \bibfield  {author} {\bibinfo {author} {\bibfnamefont {B.}~\bibnamefont {Qi}},\ }\bibfield  {title} {{\selectlanguage {en}\bibinfo {title} {Simultaneous classical communication and quantum key distribution using continuous variables}},\ }\href {https://doi.org/10.1103/PhysRevA.94.042340} {\bibfield  {journal} {\bibinfo  {journal} {Physical Review A}\ }\textbf {\bibinfo {volume} {94}},\ \bibinfo {pages} {042340} (\bibinfo {year} {2016})}\BibitemShut {NoStop}%
\bibitem [{\citenamefont {Qi}\ and\ \citenamefont {Lim}(2018)}]{qi_noise_2018}%
  \BibitemOpen
  \bibfield  {author} {\bibinfo {author} {\bibfnamefont {B.}~\bibnamefont {Qi}}\ and\ \bibinfo {author} {\bibfnamefont {C.~C.~W.}\ \bibnamefont {Lim}},\ }\bibfield  {title} {{\selectlanguage {en}\bibinfo {title} {Noise {Analysis} of {Simultaneous} {Quantum} {Key} {Distribution} and {Classical} {Communication} {Scheme} {Using} a {True} {Local} {Oscillator}}},\ }\href {https://doi.org/10.1103/PhysRevApplied.9.054008} {\bibfield  {journal} {\bibinfo  {journal} {Physical Review Applied}\ }\textbf {\bibinfo {volume} {9}},\ \bibinfo {pages} {054008} (\bibinfo {year} {2018})}\BibitemShut {NoStop}%
\bibitem [{\citenamefont {Kumar}\ \emph {et~al.}(2019)\citenamefont {Kumar}, \citenamefont {Wonfor}, \citenamefont {Penty}, \citenamefont {Spiller},\ and\ \citenamefont {White}}]{kumar_experimental_2019}%
  \BibitemOpen
  \bibfield  {author} {\bibinfo {author} {\bibfnamefont {R.}~\bibnamefont {Kumar}}, \bibinfo {author} {\bibfnamefont {A.}~\bibnamefont {Wonfor}}, \bibinfo {author} {\bibfnamefont {R.}~\bibnamefont {Penty}}, \bibinfo {author} {\bibfnamefont {T.}~\bibnamefont {Spiller}},\ and\ \bibinfo {author} {\bibfnamefont {I.}~\bibnamefont {White}},\ }\bibfield  {title} {{\selectlanguage {en}\bibinfo {title} {Experimental demonstration of single-shot quantum and classical signal transmission on single wavelength optical pulse}},\ }\href {https://doi.org/10.1038/s41598-019-47699-z} {\bibfield  {journal} {\bibinfo  {journal} {Scientific Reports}\ }\textbf {\bibinfo {volume} {9}},\ \bibinfo {pages} {11190} (\bibinfo {year} {2019})}\BibitemShut {NoStop}%
\bibitem [{\citenamefont {Zaunders}\ \emph {et~al.}(2025)\citenamefont {Zaunders}, \citenamefont {Wang}, \citenamefont {Malaney}, \citenamefont {Aguinaldo},\ and\ \citenamefont {Ralph}}]{zaunders_enhanced_2025}%
  \BibitemOpen
  \bibfield  {author} {\bibinfo {author} {\bibfnamefont {N.}~\bibnamefont {Zaunders}}, \bibinfo {author} {\bibfnamefont {Z.}~\bibnamefont {Wang}}, \bibinfo {author} {\bibfnamefont {R.}~\bibnamefont {Malaney}}, \bibinfo {author} {\bibfnamefont {R.}~\bibnamefont {Aguinaldo}},\ and\ \bibinfo {author} {\bibfnamefont {T.~C.}\ \bibnamefont {Ralph}},\ }\bibfield  {title} {{\selectlanguage {en}\bibinfo {title} {Enhanced simultaneous quantum-classical communications under composable security}},\ }\href {https://doi.org/10.1103/p34s-5217} {\bibfield  {journal} {\bibinfo  {journal} {Physical Review A}\ }\textbf {\bibinfo {volume} {112}},\ \bibinfo {pages} {042608} (\bibinfo {year} {2025})}\BibitemShut {NoStop}%
\bibitem [{\citenamefont {Nguyen}\ \emph {et~al.}(2019)\citenamefont {Nguyen}, \citenamefont {Sukachev}, \citenamefont {Bhaskar}, \citenamefont {Machielse}, \citenamefont {Levonian}, \citenamefont {Knall}, \citenamefont {Stroganov}, \citenamefont {Riedinger}, \citenamefont {Park}, \citenamefont {Lončar},\ and\ \citenamefont {Lukin}}]{nguyen_quantum_2019}%
  \BibitemOpen
  \bibfield  {author} {\bibinfo {author} {\bibfnamefont {C.}~\bibnamefont {Nguyen}}, \bibinfo {author} {\bibfnamefont {D.}~\bibnamefont {Sukachev}}, \bibinfo {author} {\bibfnamefont {M.}~\bibnamefont {Bhaskar}}, \bibinfo {author} {\bibfnamefont {B.}~\bibnamefont {Machielse}}, \bibinfo {author} {\bibfnamefont {D.}~\bibnamefont {Levonian}}, \bibinfo {author} {\bibfnamefont {E.}~\bibnamefont {Knall}}, \bibinfo {author} {\bibfnamefont {P.}~\bibnamefont {Stroganov}}, \bibinfo {author} {\bibfnamefont {R.}~\bibnamefont {Riedinger}}, \bibinfo {author} {\bibfnamefont {H.}~\bibnamefont {Park}}, \bibinfo {author} {\bibfnamefont {M.}~\bibnamefont {Lončar}},\ and\ \bibinfo {author} {\bibfnamefont {M.}~\bibnamefont {Lukin}},\ }\bibfield  {title} {{\selectlanguage {en}\bibinfo {title} {Quantum {Network} {Nodes} {Based} on {Diamond} {Qubits} with an {Efficient} {Nanophotonic} {Interface}}},\ }\href {https://doi.org/10.1103/PhysRevLett.123.183602} {\bibfield  {journal} {\bibinfo  {journal} {Physical Review Letters}\ }\textbf
  {\bibinfo {volume} {123}},\ \bibinfo {pages} {183602} (\bibinfo {year} {2019})}\BibitemShut {NoStop}%
\bibitem [{\citenamefont {Liu}\ \emph {et~al.}(2021)\citenamefont {Liu}, \citenamefont {Hu}, \citenamefont {Li}, \citenamefont {Li}, \citenamefont {Li}, \citenamefont {Liang}, \citenamefont {Zhou}, \citenamefont {Li},\ and\ \citenamefont {Guo}}]{liu_heralded_2021}%
  \BibitemOpen
  \bibfield  {author} {\bibinfo {author} {\bibfnamefont {X.}~\bibnamefont {Liu}}, \bibinfo {author} {\bibfnamefont {J.}~\bibnamefont {Hu}}, \bibinfo {author} {\bibfnamefont {Z.-F.}\ \bibnamefont {Li}}, \bibinfo {author} {\bibfnamefont {X.}~\bibnamefont {Li}}, \bibinfo {author} {\bibfnamefont {P.-Y.}\ \bibnamefont {Li}}, \bibinfo {author} {\bibfnamefont {P.-J.}\ \bibnamefont {Liang}}, \bibinfo {author} {\bibfnamefont {Z.-Q.}\ \bibnamefont {Zhou}}, \bibinfo {author} {\bibfnamefont {C.-F.}\ \bibnamefont {Li}},\ and\ \bibinfo {author} {\bibfnamefont {G.-C.}\ \bibnamefont {Guo}},\ }\bibfield  {title} {{\selectlanguage {en}\bibinfo {title} {Heralded entanglement distribution between two absorptive quantum memories}},\ }\href {https://doi.org/10.1038/s41586-021-03505-3} {\bibfield  {journal} {\bibinfo  {journal} {Nature}\ }\textbf {\bibinfo {volume} {594}},\ \bibinfo {pages} {41} (\bibinfo {year} {2021})}\BibitemShut {NoStop}%
\bibitem [{\citenamefont {Ortu}\ \emph {et~al.}(2022)\citenamefont {Ortu}, \citenamefont {Holzäpfel}, \citenamefont {Etesse},\ and\ \citenamefont {Afzelius}}]{ortu_storage_2022}%
  \BibitemOpen
  \bibfield  {author} {\bibinfo {author} {\bibfnamefont {A.}~\bibnamefont {Ortu}}, \bibinfo {author} {\bibfnamefont {A.}~\bibnamefont {Holzäpfel}}, \bibinfo {author} {\bibfnamefont {J.}~\bibnamefont {Etesse}},\ and\ \bibinfo {author} {\bibfnamefont {M.}~\bibnamefont {Afzelius}},\ }\bibfield  {title} {{\selectlanguage {en}\bibinfo {title} {Storage of photonic time-bin qubits for up to 20 ms in a rare-earth doped crystal}},\ }\href {https://doi.org/10.1038/s41534-022-00541-3} {\bibfield  {journal} {\bibinfo  {journal} {npj Quantum Information}\ }\textbf {\bibinfo {volume} {8}},\ \bibinfo {pages} {29} (\bibinfo {year} {2022})}\BibitemShut {NoStop}%
\bibitem [{\citenamefont {Braunstein}\ and\ \citenamefont {Van~Loock}(2005)}]{braunstein_quantum_2005}%
  \BibitemOpen
  \bibfield  {author} {\bibinfo {author} {\bibfnamefont {S.~L.}\ \bibnamefont {Braunstein}}\ and\ \bibinfo {author} {\bibfnamefont {P.}~\bibnamefont {Van~Loock}},\ }\bibfield  {title} {{\selectlanguage {en}\bibinfo {title} {Quantum information with continuous variables}},\ }\href {https://doi.org/10.1103/revmodphys.77.513} {\bibfield  {journal} {\bibinfo  {journal} {Reviews of Modern Physics}\ }\textbf {\bibinfo {volume} {77}},\ \bibinfo {pages} {513} (\bibinfo {year} {2005})}\BibitemShut {NoStop}%
\bibitem [{\citenamefont {García-Patrón}\ \emph {et~al.}(2009)\citenamefont {García-Patrón}, \citenamefont {Pirandola}, \citenamefont {Lloyd},\ and\ \citenamefont {Shapiro}}]{garcia-patron_reverse_2009}%
  \BibitemOpen
  \bibfield  {author} {\bibinfo {author} {\bibfnamefont {R.}~\bibnamefont {García-Patrón}}, \bibinfo {author} {\bibfnamefont {S.}~\bibnamefont {Pirandola}}, \bibinfo {author} {\bibfnamefont {S.}~\bibnamefont {Lloyd}},\ and\ \bibinfo {author} {\bibfnamefont {J.~H.}\ \bibnamefont {Shapiro}},\ }\bibfield  {title} {{\selectlanguage {en}\bibinfo {title} {Reverse {Coherent} {Information}}},\ }\href {https://doi.org/10.1103/PhysRevLett.102.210501} {\bibfield  {journal} {\bibinfo  {journal} {Physical Review Letters}\ }\textbf {\bibinfo {volume} {102}},\ \bibinfo {pages} {210501} (\bibinfo {year} {2009})}\BibitemShut {NoStop}%
\bibitem [{\citenamefont {Weedbrook}\ \emph {et~al.}(2012)\citenamefont {Weedbrook}, \citenamefont {Pirandola}, \citenamefont {García-Patrón}, \citenamefont {Cerf}, \citenamefont {Ralph}, \citenamefont {Shapiro},\ and\ \citenamefont {Lloyd}}]{weedbrook_gaussian_2012}%
  \BibitemOpen
  \bibfield  {author} {\bibinfo {author} {\bibfnamefont {C.}~\bibnamefont {Weedbrook}}, \bibinfo {author} {\bibfnamefont {S.}~\bibnamefont {Pirandola}}, \bibinfo {author} {\bibfnamefont {R.}~\bibnamefont {García-Patrón}}, \bibinfo {author} {\bibfnamefont {N.~J.}\ \bibnamefont {Cerf}}, \bibinfo {author} {\bibfnamefont {T.~C.}\ \bibnamefont {Ralph}}, \bibinfo {author} {\bibfnamefont {J.~H.}\ \bibnamefont {Shapiro}},\ and\ \bibinfo {author} {\bibfnamefont {S.}~\bibnamefont {Lloyd}},\ }\bibfield  {title} {{\selectlanguage {en}\bibinfo {title} {Gaussian quantum information}},\ }\href {https://doi.org/10.1103/RevModPhys.84.621} {\bibfield  {journal} {\bibinfo  {journal} {Reviews of Modern Physics}\ }\textbf {\bibinfo {volume} {84}},\ \bibinfo {pages} {621} (\bibinfo {year} {2012})}\BibitemShut {NoStop}%
\bibitem [{\citenamefont {Winnel}\ \emph {et~al.}(2020)\citenamefont {Winnel}, \citenamefont {Hosseinidehaj},\ and\ \citenamefont {Ralph}}]{winnel_generalized_2020}%
  \BibitemOpen
  \bibfield  {author} {\bibinfo {author} {\bibfnamefont {M.~S.}\ \bibnamefont {Winnel}}, \bibinfo {author} {\bibfnamefont {N.}~\bibnamefont {Hosseinidehaj}},\ and\ \bibinfo {author} {\bibfnamefont {T.~C.}\ \bibnamefont {Ralph}},\ }\bibfield  {title} {{\selectlanguage {en}\bibinfo {title} {Generalized quantum scissors for noiseless linear amplification}},\ }\href {https://doi.org/10.1103/PhysRevA.102.063715} {\bibfield  {journal} {\bibinfo  {journal} {Physical Review A}\ }\textbf {\bibinfo {volume} {102}},\ \bibinfo {pages} {063715} (\bibinfo {year} {2020})}\BibitemShut {NoStop}%
\bibitem [{\citenamefont {Pirandola}\ \emph {et~al.}(2020)\citenamefont {Pirandola}, \citenamefont {Andersen}, \citenamefont {Banchi}, \citenamefont {Berta}, \citenamefont {Bunandar}, \citenamefont {Colbeck}, \citenamefont {Englund}, \citenamefont {Gehring}, \citenamefont {Lupo}, \citenamefont {Ottaviani}, \citenamefont {Pereira}, \citenamefont {Razavi}, \citenamefont {Shamsul~Shaari}, \citenamefont {Tomamichel}, \citenamefont {Usenko}, \citenamefont {Vallone}, \citenamefont {Villoresi},\ and\ \citenamefont {Wallden}}]{pirandola_advances_2020}%
  \BibitemOpen
  \bibfield  {author} {\bibinfo {author} {\bibfnamefont {S.}~\bibnamefont {Pirandola}}, \bibinfo {author} {\bibfnamefont {U.~L.}\ \bibnamefont {Andersen}}, \bibinfo {author} {\bibfnamefont {L.}~\bibnamefont {Banchi}}, \bibinfo {author} {\bibfnamefont {M.}~\bibnamefont {Berta}}, \bibinfo {author} {\bibfnamefont {D.}~\bibnamefont {Bunandar}}, \bibinfo {author} {\bibfnamefont {R.}~\bibnamefont {Colbeck}}, \bibinfo {author} {\bibfnamefont {D.}~\bibnamefont {Englund}}, \bibinfo {author} {\bibfnamefont {T.}~\bibnamefont {Gehring}}, \bibinfo {author} {\bibfnamefont {C.}~\bibnamefont {Lupo}}, \bibinfo {author} {\bibfnamefont {C.}~\bibnamefont {Ottaviani}}, \bibinfo {author} {\bibfnamefont {J.~L.}\ \bibnamefont {Pereira}}, \bibinfo {author} {\bibfnamefont {M.}~\bibnamefont {Razavi}}, \bibinfo {author} {\bibfnamefont {J.}~\bibnamefont {Shamsul~Shaari}}, \bibinfo {author} {\bibfnamefont {M.}~\bibnamefont {Tomamichel}}, \bibinfo {author} {\bibfnamefont {V.~C.}\ \bibnamefont {Usenko}}, \bibinfo {author} {\bibfnamefont
  {G.}~\bibnamefont {Vallone}}, \bibinfo {author} {\bibfnamefont {P.}~\bibnamefont {Villoresi}},\ and\ \bibinfo {author} {\bibfnamefont {P.}~\bibnamefont {Wallden}},\ }\bibfield  {title} {{\selectlanguage {en}\bibinfo {title} {Advances in quantum cryptography}},\ }\href {https://doi.org/10.1364/AOP.361502} {\bibfield  {journal} {\bibinfo  {journal} {Advances in Optics and Photonics}\ }\textbf {\bibinfo {volume} {12}},\ \bibinfo {pages} {1012} (\bibinfo {year} {2020})}\BibitemShut {NoStop}%
\bibitem [{\citenamefont {Bennett}\ \emph {et~al.}(1996{\natexlab{b}})\citenamefont {Bennett}, \citenamefont {DiVincenzo}, \citenamefont {Smolin},\ and\ \citenamefont {Wootters}}]{bennett_mixed-state_1996}%
  \BibitemOpen
  \bibfield  {author} {\bibinfo {author} {\bibfnamefont {C.~H.}\ \bibnamefont {Bennett}}, \bibinfo {author} {\bibfnamefont {D.~P.}\ \bibnamefont {DiVincenzo}}, \bibinfo {author} {\bibfnamefont {J.~A.}\ \bibnamefont {Smolin}},\ and\ \bibinfo {author} {\bibfnamefont {W.~K.}\ \bibnamefont {Wootters}},\ }\bibfield  {title} {{\selectlanguage {en}\bibinfo {title} {Mixed-state entanglement and quantum error correction}},\ }\href {https://doi.org/10.1103/PhysRevA.54.3824} {\bibfield  {journal} {\bibinfo  {journal} {Physical Review A}\ }\textbf {\bibinfo {volume} {54}},\ \bibinfo {pages} {3824} (\bibinfo {year} {1996}{\natexlab{b}})}\BibitemShut {NoStop}%
\bibitem [{\citenamefont {Wolf}\ \emph {et~al.}(2004)\citenamefont {Wolf}, \citenamefont {Giedke}, \citenamefont {Krüger}, \citenamefont {Werner},\ and\ \citenamefont {Cirac}}]{wolf_gaussian_2004}%
  \BibitemOpen
  \bibfield  {author} {\bibinfo {author} {\bibfnamefont {M.~M.}\ \bibnamefont {Wolf}}, \bibinfo {author} {\bibfnamefont {G.}~\bibnamefont {Giedke}}, \bibinfo {author} {\bibfnamefont {O.}~\bibnamefont {Krüger}}, \bibinfo {author} {\bibfnamefont {R.~F.}\ \bibnamefont {Werner}},\ and\ \bibinfo {author} {\bibfnamefont {J.~I.}\ \bibnamefont {Cirac}},\ }\bibfield  {title} {{\selectlanguage {en}\bibinfo {title} {Gaussian entanglement of formation}},\ }\href {https://doi.org/10.1103/PhysRevA.69.052320} {\bibfield  {journal} {\bibinfo  {journal} {Physical Review A}\ }\textbf {\bibinfo {volume} {69}},\ \bibinfo {pages} {052320} (\bibinfo {year} {2004})}\BibitemShut {NoStop}%
\bibitem [{\citenamefont {Wolf}\ \emph {et~al.}(2006)\citenamefont {Wolf}, \citenamefont {Giedke},\ and\ \citenamefont {Cirac}}]{wolf_extremality_2006}%
  \BibitemOpen
  \bibfield  {author} {\bibinfo {author} {\bibfnamefont {M.~M.}\ \bibnamefont {Wolf}}, \bibinfo {author} {\bibfnamefont {G.}~\bibnamefont {Giedke}},\ and\ \bibinfo {author} {\bibfnamefont {J.~I.}\ \bibnamefont {Cirac}},\ }\bibfield  {title} {{\selectlanguage {en}\bibinfo {title} {Extremality of {Gaussian} {Quantum} {States}}},\ }\href {https://doi.org/10.1103/PhysRevLett.96.080502} {\bibfield  {journal} {\bibinfo  {journal} {Physical Review Letters}\ }\textbf {\bibinfo {volume} {96}},\ \bibinfo {pages} {080502} (\bibinfo {year} {2006})}\BibitemShut {NoStop}%
\bibitem [{\citenamefont {Tserkis}\ and\ \citenamefont {Ralph}(2017)}]{tserkis_quantifying_2017}%
  \BibitemOpen
  \bibfield  {author} {\bibinfo {author} {\bibfnamefont {S.}~\bibnamefont {Tserkis}}\ and\ \bibinfo {author} {\bibfnamefont {T.~C.}\ \bibnamefont {Ralph}},\ }\bibfield  {title} {{\selectlanguage {en}\bibinfo {title} {Quantifying entanglement in two-mode {Gaussian} states}},\ }\href {https://doi.org/10.1103/PhysRevA.96.062338} {\bibfield  {journal} {\bibinfo  {journal} {Physical Review A}\ }\textbf {\bibinfo {volume} {96}},\ \bibinfo {pages} {062338} (\bibinfo {year} {2017})}\BibitemShut {NoStop}%
\bibitem [{\citenamefont {Tserkis}\ \emph {et~al.}(2019)\citenamefont {Tserkis}, \citenamefont {Onoe},\ and\ \citenamefont {Ralph}}]{tserkis_quantifying_2019}%
  \BibitemOpen
  \bibfield  {author} {\bibinfo {author} {\bibfnamefont {S.}~\bibnamefont {Tserkis}}, \bibinfo {author} {\bibfnamefont {S.}~\bibnamefont {Onoe}},\ and\ \bibinfo {author} {\bibfnamefont {T.~C.}\ \bibnamefont {Ralph}},\ }\bibfield  {title} {{\selectlanguage {en}\bibinfo {title} {Quantifying entanglement of formation for two-mode {Gaussian} states: {Analytical} expressions for upper and lower bounds and numerical estimation of its exact value}},\ }\href {https://doi.org/10.1103/PhysRevA.99.052337} {\bibfield  {journal} {\bibinfo  {journal} {Physical Review A}\ }\textbf {\bibinfo {volume} {99}},\ \bibinfo {pages} {052337} (\bibinfo {year} {2019})}\BibitemShut {NoStop}%
\bibitem [{\citenamefont {Devetak}\ and\ \citenamefont {Winter}(2005)}]{devetak_distillation_2005}%
  \BibitemOpen
  \bibfield  {author} {\bibinfo {author} {\bibfnamefont {I.}~\bibnamefont {Devetak}}\ and\ \bibinfo {author} {\bibfnamefont {A.}~\bibnamefont {Winter}},\ }\bibfield  {title} {{\selectlanguage {en}\bibinfo {title} {Distillation of secret key and entanglement from quantum states}},\ }\href {https://doi.org/10.1098/rspa.2004.1372} {\bibfield  {journal} {\bibinfo  {journal} {Proceedings of the Royal Society A: Mathematical, Physical and Engineering Sciences}\ }\textbf {\bibinfo {volume} {461}},\ \bibinfo {pages} {207} (\bibinfo {year} {2005})}\BibitemShut {NoStop}%
\bibitem [{\citenamefont {Laudenbach}\ \emph {et~al.}(2018)\citenamefont {Laudenbach}, \citenamefont {Pacher}, \citenamefont {Fung}, \citenamefont {Poppe}, \citenamefont {Peev}, \citenamefont {Schrenk}, \citenamefont {Hentschel}, \citenamefont {Walther},\ and\ \citenamefont {Hübel}}]{laudenbach_continuousvariable_2018}%
  \BibitemOpen
  \bibfield  {author} {\bibinfo {author} {\bibfnamefont {F.}~\bibnamefont {Laudenbach}}, \bibinfo {author} {\bibfnamefont {C.}~\bibnamefont {Pacher}}, \bibinfo {author} {\bibfnamefont {C.~F.}\ \bibnamefont {Fung}}, \bibinfo {author} {\bibfnamefont {A.}~\bibnamefont {Poppe}}, \bibinfo {author} {\bibfnamefont {M.}~\bibnamefont {Peev}}, \bibinfo {author} {\bibfnamefont {B.}~\bibnamefont {Schrenk}}, \bibinfo {author} {\bibfnamefont {M.}~\bibnamefont {Hentschel}}, \bibinfo {author} {\bibfnamefont {P.}~\bibnamefont {Walther}},\ and\ \bibinfo {author} {\bibfnamefont {H.}~\bibnamefont {Hübel}},\ }\bibfield  {title} {{\selectlanguage {en}\bibinfo {title} {Continuous‐{Variable} {Quantum} {Key} {Distribution} with {Gaussian} {Modulation}—{The} {Theory} of {Practical} {Implementations}}},\ }\href {https://doi.org/10.1002/qute.201800011} {\bibfield  {journal} {\bibinfo  {journal} {Advanced Quantum Technologies}\ }\textbf {\bibinfo {volume} {1}},\ \bibinfo {pages} {1800011} (\bibinfo {year} {2018})}\BibitemShut
  {NoStop}%
\bibitem [{\citenamefont {Vahlbruch}\ \emph {et~al.}(2016)\citenamefont {Vahlbruch}, \citenamefont {Mehmet}, \citenamefont {Danzmann},\ and\ \citenamefont {Schnabel}}]{vahlbruch_detection_2016}%
  \BibitemOpen
  \bibfield  {author} {\bibinfo {author} {\bibfnamefont {H.}~\bibnamefont {Vahlbruch}}, \bibinfo {author} {\bibfnamefont {M.}~\bibnamefont {Mehmet}}, \bibinfo {author} {\bibfnamefont {K.}~\bibnamefont {Danzmann}},\ and\ \bibinfo {author} {\bibfnamefont {R.}~\bibnamefont {Schnabel}},\ }\bibfield  {title} {{\selectlanguage {en}\bibinfo {title} {Detection of 15 {dB} {Squeezed} {States} of {Light} and their {Application} for the {Absolute} {Calibration} of {Photoelectric} {Quantum} {Efficiency}}},\ }\href {https://doi.org/10.1103/PhysRevLett.117.110801} {\bibfield  {journal} {\bibinfo  {journal} {Physical Review Letters}\ }\textbf {\bibinfo {volume} {117}},\ \bibinfo {pages} {110801} (\bibinfo {year} {2016})}\BibitemShut {NoStop}%
\bibitem [{\citenamefont {Pirandola}\ \emph {et~al.}(2017)\citenamefont {Pirandola}, \citenamefont {Laurenza}, \citenamefont {Ottaviani},\ and\ \citenamefont {Banchi}}]{pirandola_fundamental_2017}%
  \BibitemOpen
  \bibfield  {author} {\bibinfo {author} {\bibfnamefont {S.}~\bibnamefont {Pirandola}}, \bibinfo {author} {\bibfnamefont {R.}~\bibnamefont {Laurenza}}, \bibinfo {author} {\bibfnamefont {C.}~\bibnamefont {Ottaviani}},\ and\ \bibinfo {author} {\bibfnamefont {L.}~\bibnamefont {Banchi}},\ }\bibfield  {title} {{\selectlanguage {en}\bibinfo {title} {Fundamental limits of repeaterless quantum communications}},\ }\href {https://doi.org/10.1038/ncomms15043} {\bibfield  {journal} {\bibinfo  {journal} {Nature Communications}\ }\textbf {\bibinfo {volume} {8}},\ \bibinfo {pages} {15043} (\bibinfo {year} {2017})}\BibitemShut {NoStop}%
\bibitem [{\citenamefont {Guanzon}\ \emph {et~al.}(2022)\citenamefont {Guanzon}, \citenamefont {Winnel}, \citenamefont {Lund},\ and\ \citenamefont {Ralph}}]{guanzon_ideal_2022}%
  \BibitemOpen
  \bibfield  {author} {\bibinfo {author} {\bibfnamefont {J.~J.}\ \bibnamefont {Guanzon}}, \bibinfo {author} {\bibfnamefont {M.~S.}\ \bibnamefont {Winnel}}, \bibinfo {author} {\bibfnamefont {A.~P.}\ \bibnamefont {Lund}},\ and\ \bibinfo {author} {\bibfnamefont {T.~C.}\ \bibnamefont {Ralph}},\ }\bibfield  {title} {{\selectlanguage {en}\bibinfo {title} {Ideal {Quantum} {Teleamplification} up to a {Selected} {Energy} {Cutoff} {Using} {Linear} {Optics}}},\ }\href {https://doi.org/10.1103/PhysRevLett.128.160501} {\bibfield  {journal} {\bibinfo  {journal} {Physical Review Letters}\ }\textbf {\bibinfo {volume} {128}},\ \bibinfo {pages} {160501} (\bibinfo {year} {2022})}\BibitemShut {NoStop}%
\bibitem [{\citenamefont {Dias}\ \emph {et~al.}(2020)\citenamefont {Dias}, \citenamefont {Winnel}, \citenamefont {Hosseinidehaj},\ and\ \citenamefont {Ralph}}]{dias_quantum_2020}%
  \BibitemOpen
  \bibfield  {author} {\bibinfo {author} {\bibfnamefont {J.}~\bibnamefont {Dias}}, \bibinfo {author} {\bibfnamefont {M.~S.}\ \bibnamefont {Winnel}}, \bibinfo {author} {\bibfnamefont {N.}~\bibnamefont {Hosseinidehaj}},\ and\ \bibinfo {author} {\bibfnamefont {T.~C.}\ \bibnamefont {Ralph}},\ }\bibfield  {title} {{\selectlanguage {en}\bibinfo {title} {Quantum repeater for continuous-variable entanglement distribution}},\ }\href {https://doi.org/10.1103/PhysRevA.102.052425} {\bibfield  {journal} {\bibinfo  {journal} {Physical Review A}\ }\textbf {\bibinfo {volume} {102}},\ \bibinfo {pages} {052425} (\bibinfo {year} {2020})}\BibitemShut {NoStop}%
\bibitem [{\citenamefont {Azuma}\ \emph {et~al.}(2023)\citenamefont {Azuma}, \citenamefont {Economou}, \citenamefont {Elkouss}, \citenamefont {Hilaire}, \citenamefont {Jiang}, \citenamefont {Lo},\ and\ \citenamefont {Tzitrin}}]{azuma_quantum_2023}%
  \BibitemOpen
  \bibfield  {author} {\bibinfo {author} {\bibfnamefont {K.}~\bibnamefont {Azuma}}, \bibinfo {author} {\bibfnamefont {S.~E.}\ \bibnamefont {Economou}}, \bibinfo {author} {\bibfnamefont {D.}~\bibnamefont {Elkouss}}, \bibinfo {author} {\bibfnamefont {P.}~\bibnamefont {Hilaire}}, \bibinfo {author} {\bibfnamefont {L.}~\bibnamefont {Jiang}}, \bibinfo {author} {\bibfnamefont {H.-K.}\ \bibnamefont {Lo}},\ and\ \bibinfo {author} {\bibfnamefont {I.}~\bibnamefont {Tzitrin}},\ }\bibfield  {title} {{\selectlanguage {en}\bibinfo {title} {Quantum repeaters: {From} quantum networks to the quantum internet}},\ }\href {https://doi.org/10.1103/RevModPhys.95.045006} {\bibfield  {journal} {\bibinfo  {journal} {Reviews of Modern Physics}\ }\textbf {\bibinfo {volume} {95}},\ \bibinfo {pages} {045006} (\bibinfo {year} {2023})}\BibitemShut {NoStop}%
\bibitem [{\citenamefont {Ralph}\ and\ \citenamefont {Lund}(2009)}]{ralph_nondeterministic_2009}%
  \BibitemOpen
  \bibfield  {author} {\bibinfo {author} {\bibfnamefont {T.~C.}\ \bibnamefont {Ralph}}\ and\ \bibinfo {author} {\bibfnamefont {A.~P.}\ \bibnamefont {Lund}},\ }\bibfield  {title} {{\selectlanguage {en}\bibinfo {title} {Nondeterministic {Noiseless} {Linear} {Amplification} of {Quantum} {Systems}}},\ }in\ \href {https://doi.org/10.1063/1.3131295} {{\selectlanguage {en}\emph {\bibinfo {booktitle} {{AIP} {Conference} {Proceedings}}}}}\ (\bibinfo  {publisher} {AIP},\ \bibinfo {address} {Calgary (Canada)},\ \bibinfo {year} {2009})\ pp.\ \bibinfo {pages} {155--160}\BibitemShut {NoStop}%
\bibitem [{\citenamefont {Xiang}\ \emph {et~al.}(2010)\citenamefont {Xiang}, \citenamefont {Ralph}, \citenamefont {Lund}, \citenamefont {Walk},\ and\ \citenamefont {Pryde}}]{xiang_heralded_2010}%
  \BibitemOpen
  \bibfield  {author} {\bibinfo {author} {\bibfnamefont {G.~Y.}\ \bibnamefont {Xiang}}, \bibinfo {author} {\bibfnamefont {T.~C.}\ \bibnamefont {Ralph}}, \bibinfo {author} {\bibfnamefont {A.~P.}\ \bibnamefont {Lund}}, \bibinfo {author} {\bibfnamefont {N.}~\bibnamefont {Walk}},\ and\ \bibinfo {author} {\bibfnamefont {G.~J.}\ \bibnamefont {Pryde}},\ }\bibfield  {title} {{\selectlanguage {en}\bibinfo {title} {Heralded noiseless linear amplification and distillation of entanglement}},\ }\href {https://doi.org/10.1038/nphoton.2010.35} {\bibfield  {journal} {\bibinfo  {journal} {Nature Photonics}\ }\textbf {\bibinfo {volume} {4}},\ \bibinfo {pages} {316} (\bibinfo {year} {2010})}\BibitemShut {NoStop}%
\bibitem [{\citenamefont {Dias}\ \emph {et~al.}(2022)\citenamefont {Dias}, \citenamefont {Winnel}, \citenamefont {Munro}, \citenamefont {Ralph},\ and\ \citenamefont {Nemoto}}]{dias_distributing_2022}%
  \BibitemOpen
  \bibfield  {author} {\bibinfo {author} {\bibfnamefont {J.}~\bibnamefont {Dias}}, \bibinfo {author} {\bibfnamefont {M.~S.}\ \bibnamefont {Winnel}}, \bibinfo {author} {\bibfnamefont {W.~J.}\ \bibnamefont {Munro}}, \bibinfo {author} {\bibfnamefont {T.~C.}\ \bibnamefont {Ralph}},\ and\ \bibinfo {author} {\bibfnamefont {K.}~\bibnamefont {Nemoto}},\ }\bibfield  {title} {{\selectlanguage {en}\bibinfo {title} {Distributing entanglement in first-generation discrete- and continuous-variable quantum repeaters}},\ }\href {https://doi.org/10.1103/PhysRevA.106.052604} {\bibfield  {journal} {\bibinfo  {journal} {Physical Review A}\ }\textbf {\bibinfo {volume} {106}},\ \bibinfo {pages} {052604} (\bibinfo {year} {2022})}\BibitemShut {NoStop}%
\bibitem [{\citenamefont {Rohde}\ \emph {et~al.}(2025)\citenamefont {Rohde}, \citenamefont {Huang}, \citenamefont {Ouyang}, \citenamefont {Huang}, \citenamefont {Su}, \citenamefont {Devitt}, \citenamefont {Ramakrishnan}, \citenamefont {Mantri}, \citenamefont {Tan}, \citenamefont {Liu}, \citenamefont {Harrison}, \citenamefont {Radhakrishnan}, \citenamefont {Brennen}, \citenamefont {Baragiola}, \citenamefont {Dowling}, \citenamefont {Byrnes},\ and\ \citenamefont {Munro}}]{rohde_quantum_2025}%
  \BibitemOpen
  \bibfield  {author} {\bibinfo {author} {\bibfnamefont {P.~P.}\ \bibnamefont {Rohde}}, \bibinfo {author} {\bibfnamefont {Z.}~\bibnamefont {Huang}}, \bibinfo {author} {\bibfnamefont {Y.}~\bibnamefont {Ouyang}}, \bibinfo {author} {\bibfnamefont {H.-L.}\ \bibnamefont {Huang}}, \bibinfo {author} {\bibfnamefont {Z.-E.}\ \bibnamefont {Su}}, \bibinfo {author} {\bibfnamefont {S.}~\bibnamefont {Devitt}}, \bibinfo {author} {\bibfnamefont {R.}~\bibnamefont {Ramakrishnan}}, \bibinfo {author} {\bibfnamefont {A.}~\bibnamefont {Mantri}}, \bibinfo {author} {\bibfnamefont {S.-H.}\ \bibnamefont {Tan}}, \bibinfo {author} {\bibfnamefont {N.}~\bibnamefont {Liu}}, \bibinfo {author} {\bibfnamefont {S.}~\bibnamefont {Harrison}}, \bibinfo {author} {\bibfnamefont {C.}~\bibnamefont {Radhakrishnan}}, \bibinfo {author} {\bibfnamefont {G.~K.}\ \bibnamefont {Brennen}}, \bibinfo {author} {\bibfnamefont {B.~Q.}\ \bibnamefont {Baragiola}}, \bibinfo {author} {\bibfnamefont {J.~P.}\ \bibnamefont {Dowling}}, \bibinfo {author} {\bibfnamefont
  {T.}~\bibnamefont {Byrnes}},\ and\ \bibinfo {author} {\bibfnamefont {W.~J.}\ \bibnamefont {Munro}},\ }\href {https://doi.org/10.48550/arXiv.2501.12107} {\bibinfo {title} {The {Quantum} {Internet} ({Technical} {Version})}} (\bibinfo {year} {2025}),\ \bibinfo {note} {arXiv:2501.12107 [quant-ph]}\BibitemShut {NoStop}%
\bibitem [{\citenamefont {Andersen}\ and\ \citenamefont {Ralph}(2013)}]{andersen_high-fidelity_2013}%
  \BibitemOpen
  \bibfield  {author} {\bibinfo {author} {\bibfnamefont {U.~L.}\ \bibnamefont {Andersen}}\ and\ \bibinfo {author} {\bibfnamefont {T.~C.}\ \bibnamefont {Ralph}},\ }\bibfield  {title} {{\selectlanguage {en}\bibinfo {title} {High-{Fidelity} {Teleportation} of {Continuous}-{Variable} {Quantum} {States} {Using} {Delocalized} {Single} {Photons}}},\ }\href {https://doi.org/10.1103/PhysRevLett.111.050504} {\bibfield  {journal} {\bibinfo  {journal} {Physical Review Letters}\ }\textbf {\bibinfo {volume} {111}},\ \bibinfo {pages} {050504} (\bibinfo {year} {2013})}\BibitemShut {NoStop}%
\bibitem [{\citenamefont {Eisert}\ \emph {et~al.}(2002)\citenamefont {Eisert}, \citenamefont {Scheel},\ and\ \citenamefont {Plenio}}]{eisert_distilling_2002}%
  \BibitemOpen
  \bibfield  {author} {\bibinfo {author} {\bibfnamefont {J.}~\bibnamefont {Eisert}}, \bibinfo {author} {\bibfnamefont {S.}~\bibnamefont {Scheel}},\ and\ \bibinfo {author} {\bibfnamefont {M.~B.}\ \bibnamefont {Plenio}},\ }\bibfield  {title} {{\selectlanguage {en}\bibinfo {title} {Distilling {Gaussian} {States} with {Gaussian} {Operations} is {Impossible}}},\ }\href {https://doi.org/10.1103/PhysRevLett.89.137903} {\bibfield  {journal} {\bibinfo  {journal} {Physical Review Letters}\ }\textbf {\bibinfo {volume} {89}},\ \bibinfo {pages} {137903} (\bibinfo {year} {2002})}\BibitemShut {NoStop}%
\bibitem [{\citenamefont {Olivares}(2012)}]{olivares_quantum_2012}%
  \BibitemOpen
  \bibfield  {author} {\bibinfo {author} {\bibfnamefont {S.}~\bibnamefont {Olivares}},\ }\bibfield  {title} {{\selectlanguage {en}\bibinfo {title} {Quantum optics in the phase space - {A} tutorial on {Gaussian} states}},\ }\href {https://doi.org/10.1140/epjst/e2012-01532-4} {\bibfield  {journal} {\bibinfo  {journal} {The European Physical Journal Special Topics}\ }\textbf {\bibinfo {volume} {203}},\ \bibinfo {pages} {3} (\bibinfo {year} {2012})},\ \bibinfo {note} {arXiv:1111.0786 [quant-ph]}\BibitemShut {NoStop}%
\bibitem [{\citenamefont {Curtright}\ \emph {et~al.}(2014)\citenamefont {Curtright}, \citenamefont {Fairlie},\ and\ \citenamefont {Zachos}}]{curtright_concise_2014}%
  \BibitemOpen
  \bibfield  {author} {\bibinfo {author} {\bibfnamefont {T.~L.}\ \bibnamefont {Curtright}}, \bibinfo {author} {\bibfnamefont {D.}~\bibnamefont {Fairlie}},\ and\ \bibinfo {author} {\bibfnamefont {C.}~\bibnamefont {Zachos}},\ }\href@noop {} {\emph {\bibinfo {title} {A concise treatise on quantum mechanics in phase space}}}\ (\bibinfo  {publisher} {World Scientific},\ \bibinfo {address} {Singapore Hackensack London},\ \bibinfo {year} {2014})\BibitemShut {NoStop}%
\bibitem [{\citenamefont {Ralph}\ \emph {et~al.}(1999)\citenamefont {Ralph}, \citenamefont {Lam},\ and\ \citenamefont {Polkinghorne}}]{ralph_characterizing_1999}%
  \BibitemOpen
  \bibfield  {author} {\bibinfo {author} {\bibfnamefont {T.~C.}\ \bibnamefont {Ralph}}, \bibinfo {author} {\bibfnamefont {P.~K.}\ \bibnamefont {Lam}},\ and\ \bibinfo {author} {\bibfnamefont {R.~E.~S.}\ \bibnamefont {Polkinghorne}},\ }\bibfield  {title} {\bibinfo {title} {Characterizing teleportation in optics},\ }\href {https://doi.org/10.1088/1464-4266/1/4/321} {\bibfield  {journal} {\bibinfo  {journal} {Journal of Optics B: Quantum and Semiclassical Optics}\ }\textbf {\bibinfo {volume} {1}},\ \bibinfo {pages} {483} (\bibinfo {year} {1999})}\BibitemShut {NoStop}%
\bibitem [{Note1()}]{Note1}%
  \BibitemOpen
  \bibinfo {note} {It may seem unreasonable to claim that Bob does not have a backwards communications channel, as we do in the body of the paper, and then quantify the distillable entanglement explicitly in terms of a reverse-assisted capacity. We are obliged to ignore this apparent contradiction for practical reasons, since the forward capacity is trivially known and not especially illustrative. However, practical implementations of the protocol would likely have two-way communication accessible in some form or another and so evaluating the RCI is not unreasonable.}\BibitemShut {Stop}%
\bibitem [{\citenamefont {Devetak}\ and\ \citenamefont {Shor}(2005)}]{devetak_capacity_2005}%
  \BibitemOpen
  \bibfield  {author} {\bibinfo {author} {\bibfnamefont {I.}~\bibnamefont {Devetak}}\ and\ \bibinfo {author} {\bibfnamefont {P.~W.}\ \bibnamefont {Shor}},\ }\bibfield  {title} {{\selectlanguage {en}\bibinfo {title} {The {Capacity} of a {Quantum} {Channel} for {Simultaneous} {Transmission} of {Classical} and {Quantum} {Information}}},\ }\href {https://doi.org/10.1007/s00220-005-1317-6} {\bibfield  {journal} {\bibinfo  {journal} {Communications in Mathematical Physics}\ }\textbf {\bibinfo {volume} {256}},\ \bibinfo {pages} {287} (\bibinfo {year} {2005})}\BibitemShut {NoStop}%
\bibitem [{\citenamefont {Weedbrook}\ \emph {et~al.}(2004)\citenamefont {Weedbrook}, \citenamefont {Lance}, \citenamefont {Bowen}, \citenamefont {Symul}, \citenamefont {Ralph},\ and\ \citenamefont {Lam}}]{weedbrook_quantum_2004}%
  \BibitemOpen
  \bibfield  {author} {\bibinfo {author} {\bibfnamefont {C.}~\bibnamefont {Weedbrook}}, \bibinfo {author} {\bibfnamefont {A.~M.}\ \bibnamefont {Lance}}, \bibinfo {author} {\bibfnamefont {W.~P.}\ \bibnamefont {Bowen}}, \bibinfo {author} {\bibfnamefont {T.}~\bibnamefont {Symul}}, \bibinfo {author} {\bibfnamefont {T.~C.}\ \bibnamefont {Ralph}},\ and\ \bibinfo {author} {\bibfnamefont {P.~K.}\ \bibnamefont {Lam}},\ }\bibfield  {title} {{\selectlanguage {en}\bibinfo {title} {Quantum {Cryptography} {Without} {Switching}}},\ }\href {https://doi.org/10.1103/PhysRevLett.93.170504} {\bibfield  {journal} {\bibinfo  {journal} {Physical Review Letters}\ }\textbf {\bibinfo {volume} {93}},\ \bibinfo {pages} {170504} (\bibinfo {year} {2004})}\BibitemShut {NoStop}%
\bibitem [{\citenamefont {Navascués}\ \emph {et~al.}(2006)\citenamefont {Navascués}, \citenamefont {Grosshans},\ and\ \citenamefont {Acín}}]{navascues_optimality_2006}%
  \BibitemOpen
  \bibfield  {author} {\bibinfo {author} {\bibfnamefont {M.}~\bibnamefont {Navascués}}, \bibinfo {author} {\bibfnamefont {F.}~\bibnamefont {Grosshans}},\ and\ \bibinfo {author} {\bibfnamefont {A.}~\bibnamefont {Acín}},\ }\bibfield  {title} {{\selectlanguage {en}\bibinfo {title} {Optimality of {Gaussian} {Attacks} in {Continuous}-{Variable} {Quantum} {Cryptography}}},\ }\href {https://doi.org/10.1103/PhysRevLett.97.190502} {\bibfield  {journal} {\bibinfo  {journal} {Physical Review Letters}\ }\textbf {\bibinfo {volume} {97}},\ \bibinfo {pages} {190502} (\bibinfo {year} {2006})}\BibitemShut {NoStop}%
\bibitem [{\citenamefont {García-Patrón}\ and\ \citenamefont {Cerf}(2006)}]{garcia-patron_unconditional_2006}%
  \BibitemOpen
  \bibfield  {author} {\bibinfo {author} {\bibfnamefont {R.}~\bibnamefont {García-Patrón}}\ and\ \bibinfo {author} {\bibfnamefont {N.~J.}\ \bibnamefont {Cerf}},\ }\bibfield  {title} {{\selectlanguage {en}\bibinfo {title} {Unconditional {Optimality} of {Gaussian} {Attacks} against {Continuous}-{Variable} {Quantum} {Key} {Distribution}}},\ }\href {https://doi.org/10.1103/PhysRevLett.97.190503} {\bibfield  {journal} {\bibinfo  {journal} {Physical Review Letters}\ }\textbf {\bibinfo {volume} {97}},\ \bibinfo {pages} {190503} (\bibinfo {year} {2006})}\BibitemShut {NoStop}%
\end{thebibliography}
%

\FloatBarrier
\appendix
\begin{widetext}

\section{Characterising the entangled state}
\label{app:derivation}
In this section we will derive the mean vector and covariance matrix of the ensemble state $\hat \rho_{AB}^\mathrm{out}$ in full. We begin with Alice and Bob each generating a two-mode squeezed vacuum state of squeezing $r_A$ and $r_B$ respectively; Alice and Bob label each mode of their state as a retained mode ($A$, $B$) and a sent mode ($A_S$, $B_S$) such that each state $\hat \rho_{AA_S}$ and $\hat \rho_{B_SB}$ is fully characterised by the mean and covariance matrix
\begin{align}
    \bm{\mu}_{AA_S} &= (0, 0, 0, 0)^T \\
    V_{AA_S} &= \begin{pmatrix}
        \cosh 2r_A \id & \sinh 2r_A \sz \\
        \sinh 2r_A \sz & \cosh 2r_A \id
    \end{pmatrix} \\
    \bm{\mu}_{B_SB} &= (0, 0, 0, 0)^T \\
    V_{B_SB} &= \begin{pmatrix}
        \cosh 2r_B \id & \sinh 2r_B \sz \\
        \sinh 2r_B \sz & \cosh 2r_B \id
    \end{pmatrix}.
\end{align}
We consider first the simplified case where Alice modulates her signal by some fixed and arbitrary displacement, which we label $\bm{d} = (d_x, d_y)$, and then specify our results to the case where the displacement is randomly chosen from some classical alphabet.

The next three steps in the protocol are the classical modulation of Alice's outgoing mode $A_S$, transmission of $A_S$ across a lossy and noisy channel $\mathcal{E}$, and mixing of the two modes $A_S$ and $B_S$ on a balanced beamsplitter prior to teleportation. The first operation is Alice's displacement of her outgoing mode $A_S$ such that
\begin{align}
    \bm{\mu}_{AA_S}^d \longrightarrow (0, 0, d_x, d_y)^T.
\end{align}
The next operation is transmission of the mode $A_S$ via the channel $\mathcal{E}$. This is modelled as a beamsplitter of transmissivity $\eta$ corresponding to the loss of the channel, with the other input port of the beamsplitter accepting a thermal state of mean photon number parameter $\overline{n} = \eta \e / 2(1-\eta)$; this enacts the map
\begin{align}
    V_{AA_S}^d \rightarrow \begin{pmatrix}
        \cosh(2r_A) \id & \sqrt{\eta}\sinh(2r_A) \sz \\
        \sqrt{\eta} \sinh(2r_A) \sz & \left[ \eta(\cosh 2r_A + \e - 1) + 1 \right] \id
    \end{pmatrix} \equiv \begin{pmatrix}
        a_{00} \id & c_{00} \sz \\
        c_{00} \sz & b_{00} \id
    \end{pmatrix}.
\end{align}
At this stage the joint state $\hat \rho_{AA_SB_SB}^d$ is a separable Gaussian state, and so has trivial mean and covariance:
\begin{align}
    \bm{\mu}_{AA_SB_SB}^d &= (0,0,\sqrt{\eta} d_x,\sqrt{\eta} d_y,0,0,0,0)^T \\
    V_{AA_SB_SB}^d &= V_{AA_S}^d \oplus V_{B_SB}.
\end{align}
After transmission to Bob, the two `sent' modes $A_S$ and $B_S$ are entangled. This is done on a balanced beamsplitter, with appropriate symplectic representation
\begin{align}
    S_{A_SB_S} &= \begin{pmatrix}
        \id & 0 & 0 & 0 \\
        0 & \frac{1}{\sqrt{2}} \id & -\frac{1}{\sqrt{2}} \id & 0 \\
        0 & \frac{1}{\sqrt{2}} \id & \frac{1}{\sqrt{2}} \id & 0 \\
        0 & 0 & 0 & \id
    \end{pmatrix}
\end{align}
such that the joint output state $(\hat \rho_{AA_SB_SB}^d)'$ is a four-mode mixed entangled Gaussian state with mean and covariance
\begin{align}
    (\bm{\mu}_{AA_SB_SB}^d)' &= S_{A_SB_S} \cdot \bm{\mu}_{AA_SB_SB} \\
    &= (0,0, \frac{\sqrt{\eta}d_x}{\sqrt{2}}, \frac{\sqrt{\eta}d_y}{\sqrt{2}}, 0, 0, 0,0) \\
    &\equiv (0,0, \alpha_x, \alpha_y, 0, 0, 0,0) \\
    (V_{AA_SB_SB}^d)' &= S_{A_SB_S} \cdot V_{AA_SB_SB} \cdot S_{A_SB_S}^T \\
    &= \begin{pmatrix}
            a \id & \frac{c}{\sqrt{2}} \sz & \frac{c}{\sqrt{2}} \sz & 0 \\
            \frac{c}{\sqrt{2}} \sz & \frac{b+\cosh2r_B}{2} \id & \frac{b-\cosh2r_B}{2} \id & -\frac{\sinh 2r_B}{\sqrt{2}} \ \sz \\
            \frac{c}{\sqrt{2}} \sz & \frac{b-\cosh 2r_B}{2} \id & \frac{b+\cosh 2r_B}{2} \id & \frac{\sinh 2r_B}{\sqrt{2}} \ \sz \\
            0 & -\frac{\sinh 2r_B}{\sqrt{2}} \ \sz & \frac{\sinh 2r_B}{\sqrt{2}} \ \sz & \cosh 2r_B \id
        \end{pmatrix}.
\end{align}
In order to perform the teleportation, Bob first performs a dual homodyne measurement on the entangled modes $A_S$, $B_S$, measuring the $\hat q$ quadrature on the mode $B_S$ and the $\hat p$ quadrature on the mode $A_S$. Homodyne measurement of a field quadrature is defined as equivalent to projection onto the quadrature eigenstate \cite{eisert_distilling_2002}; the measurement therefore has the equivalent projector
\begin{align}
    \hat \Pi &= \id \otimes \ket{\hat p = y_0}_{A_s}\bra{\hat p = y_0}_{A_s} \otimes \ket{\hat q = x_0}_{B_s}\bra{\hat q = x_0}_{B_s} \otimes \id,
\end{align}
where the single-mode projectors $\ket{\hat p = y_0}_{A_s}\bra{\hat p = y_0}_{A_s}$, $\ket{\hat q = x_0}_{B_s}\bra{\hat q = x_0}_{B_s}$ represent projection onto the $\hat p$ eigenstate of the mode $A_S$ with measurement result $y_0$ and projection onto the $\hat q$ eigenstate of the mode $B_S$ with measurement result $x_0$ respectively. The probability of obtaining a specific measurement result $\bm{x} = (x_0, y_0)^T$ for each quadrature is given by the norm of the state post-measurement,
\begin{align}
    P(x_0, y_0) &= \Tr \left[ (\hat \rho_{AA_SB_SB}^d)' \ \hat \Pi \right],
\end{align}
and the two-mode entangled output state $\hat \rho_{AB}$ is conditioned on obtaining the specific outcome $x_0, y_0$; that is,
\begin{align}
    \left[ \hat \rho_{AB}^d \right]_{x_0, y_0} &= \frac{\Tr_{A_SB_S} \left[ (\hat \rho_{AA_SB_SB}^d)' \ \hat \Pi \right]}{\Tr \left[ (\hat \rho_{AA_SB_SB}^d)' \ \hat \Pi \right]}.
\end{align}

To do this, we switch to a Wigner function formalism and follow the example of continuous-variable teleportation outlined in \cite{braunstein_quantum_2005}. The Wigner function of the Gaussian state $(\hat \rho_{AA_SB_SB}^d)'$ is given by \cite{weedbrook_gaussian_2012}
\begin{align}
    W_{AA_SB_SB}^d(\bm{q}) &= \frac{\exp \left[-\frac{1}{2} (\bm{q} - \bm{\mu}_{AA_SB_SB}^d)^T \cdot (V_{AA_SB_SB}^d)^{-1} \cdot (\bm{q} - \bm{\mu}_{AA_SB_SB}^d) \right]}{(2\pi)^N \sqrt{\det (V_{AA_SB_SB}^d)}}
\end{align}
for the real-valued phase-space position vector $\bm{q} = (q_A,p_A,q_{A_S}, \dots)^T$, $\bm{q} \in \mathbb{R}^{8}$, corresponding to the $4$ pairs of quadrature operators $\bm{\hat q} = (\hat q_A, \hat p_A, \hat q_{A_S}, \dots)$ for the state of $N = 4$ modes. Quantum measurements in the Wigner formalism are given by the weighted integral of the state function with the Wigner function of the measurement operator over the quadratures corresponding to the measured modes \cite{olivares_quantum_2012}; thus, the (unnormalised) Wigner function of the state $\left[ \hat \rho_{AB}^d \right]_{x_0, y_0}$ is given explicitly by
\begin{align}
    \left[ W_{AB}^d \right]_{x_0,y_0}(q_A, p_A, q_B, p_B) &= (4\pi)^2 \int \dd q_{A_S} \ \dd p_{A_S} \ \dd q_{B_S} \ \dd p_{B_S} \ W_{AA_SB_SB}(\bm{x}) \ \frac{\delta(q_{B_S} - x_0)}{4\pi} \frac{\delta(p_{A_S} - y_0)}{4\pi},
\end{align}
where we use the known result that the Wigner function of a position or momentum eigenstate of the quadrature $\hat x$, for a specific measurement result $X$, is a delta function in the phase-space variable $x$ located at $X$ \cite{curtright_concise_2014}. The norm of the state is given by the integral of the Wigner function over all modes
\begin{align}
    P(x_0,y_0) &= \int \dd^4 \bm{q} \ \left[ W_{AB}^d \right]_{x_0,y_0}(q_A, p_A, q_B, p_B).
\end{align}
Evaluating the above explicitly, we find
\begin{align}
    P(x_0, y_0) &= \frac{1}{\pi}\frac{\exp \left(-\frac{1}{2}\frac{(\sqrt{2}x_0-\alpha_x)^2+\sqrt{2}y_0-\alpha_y)^2}{b+\cosh (2 r_B)}\right)}{b+\cosh(2r_B)}.
\end{align}
It is instructive now to consider the state represented by the function $\left[ W_{AB} \right]_{x_0,y_0}$, which we recall specifically corresponds to the single two-mode entangled state shared by Alice and Bob assuming an initial classical displacement $\bm{d} = (d_x, d_y)^T$ and conditioned on receiving the homodyne measurement results $x_0$ and $y_0$. It is known that CV teleportation introduces fluctuations on the displacement of the teleported modes, proportional to the outcomes obtained during measurement and varying shot-by-shot \cite{eisert_distilling_2002, braunstein_quantum_2005}. We can calculate these fluctuations directly by looking at the mean of the state $\left[ W_{AB} \right]_{x_0,y_0}$, which we calculate via the statistical moments of the Wigner function according to \cite{olivares_quantum_2012}
\begin{align} \label{eq:wignerfunc_moments}
    \Tr [\hat q_i \hat \rho] &= \int \dd^N \bm{q} \ q_i \ W_{\hat \rho}(\bm{q}).
\end{align}
We determine that
\begin{align}
    \left[ \bm{\mu}_{AB}^d \right]_{x_0,y_0} &= \left( \frac{c(\sqrt{2}x_0-\alpha_x)}{b+\cosh(2r_B)}, \frac{-c(\sqrt{2}y_0-\alpha_y)}{b+\cosh(2r_B)}, \frac{\sinh2r_B(\sqrt{2}x_0-\alpha_x)}{b+\cosh(2r_B)}, \frac{\sinh2r_B(\sqrt{2}y_0-\alpha_y)}{b+\cosh(2r_B)} \right)^T. \label{eq:singleshot_mean}
\end{align}
To propagate the sent mode $A_S$ onto the remaining EPR mode $B$, Bob applies the displacement
\begin{align}
    \hat D_B( -g_x\bm{x} )
\end{align}
The gain factor $g_x$ can be freely chosen by Bob; we assume $g_x = -\sqrt{2}\tanh r_B \bm{x}$ such that the teleporter becomes an effective pure-loss channel in the ensemble limit \cite{ralph_characterizing_1999}, with transmissivity $\tau = \tanh^2 r_B$.

Removing the displacement arising from the classical modulation is more involved. Since Bob does not have direct access to the actual classical modulation $\alpha_{x,y}$, he can only estimate the correct displacement based off his measurement results $x_0, y_0$, which from $P(x_0,y_0)$ he knows are Gaussian random variables centred around the classical modulation values. We therefore use the threshold discrimination algorithm described in \cite{zaunders_enhanced_2025} to define an estimate $\langle \bm \alpha \rangle$ of the true classical displacement based off the measurement results $x_0,y_0$:
\begin{align}
    \langle \bm \alpha \rangle = \begin{cases}
        (\alpha_x ,  \alpha_y) & x_0 \geq 0, y_0 \geq 0 \\
        (\alpha_x , -\alpha_y) & x_0 \geq 0, y_0 < 0 \\
        (-\alpha_x,  \alpha_y) & x_0 < 0, y_0 \geq 0 \\
        (-\alpha_x, -\alpha_y) & x_0 < 0, y_0 < 0.
    \end{cases}
\end{align}
That is, if Bob measures $x_0,y_0 \geq 0$, he correctly assumes Alice has sent the symbol $\bm d = (d_x, d_y)^T$; if he measures $x_0 \geq 0$, $y_0 < 0$, he erroneously assumes Alice has sent the symbol $\bm d = (d_x, -d_y)^T$, etc. Based on this, the displacement operation Bob performs to return the signal to zero-mean changes dependent on his measurements. We define this conditional displacement as
\begin{align}
    \hat D(-g_\alpha\langle \bm \alpha \rangle) &= \begin{cases}
        D\left[-g_\alpha(\alpha_x ,  \alpha_y)^T \right], & x_0 \geq 0, y_0 \geq 0 \\
        D\left[-g_\alpha(\alpha_x , -\alpha_y)^T \right], & x_0 \geq 0, y_0 < 0 \\
        D\left[-g_\alpha(-\alpha_x,  \alpha_y)^T \right], & x_0 < 0, y_0 \geq 0 \\
        D\left[-g_\alpha(-\alpha_x, -\alpha_y)^T \right], & x_0 < 0, y_0 < 0
    \end{cases}
\end{align}
where we set $g_\alpha = \sqrt{\tau \eta}$ to account for the attenuation of the signal from both the channel $\mathcal{E}$ and the effective teleporter attenuation $\tau$. After both displacements, the mean of the output state becomes
\begin{align}
    \left[ \bm{\mu}_{AB}^d \right]_{x_0,y_0} &\longrightarrow
    \begin{cases}
        (\mu_{q_A}, \mu_{p_A}, \mu_{q_B}^+, \mu_{p_B}^+)^T & x_0 \geq 0, y_0 \geq 0 \\
        (\mu_{q_A}, \mu_{p_A}, \mu_{q_B}^+, \mu_{p_B}^-)^T & x_0 \geq 0, y_0 \geq 0 \\
        (\mu_{q_A}, \mu_{p_A}, \mu_{q_B}^-, \mu_{p_B}^+)^T & x_0 \geq 0, y_0 \geq 0 \\
        (\mu_{q_A}, \mu_{p_A}, \mu_{q_B}^-, \mu_{p_B}^-)^T & x_0 \geq 0, y_0 \geq 0 \\
    \end{cases}.
\end{align}
for $\mu_{q_A} = \frac{c(\sqrt{2}x_0-\alpha_x)}{b+\cosh(2r_B)}$, $\mu_{p_A} = \frac{c(\sqrt{2}y_0-\alpha_y)}{b+\cosh(2r_B)}$ and
\begin{align}
    \mu_{q_B}^+ &= \frac{(1-b)(\sqrt{2}x_0-\alpha_x)\tanh r_B}{b+\cosh(2r_B)} \\
    \mu_{p_B}^+ &= \frac{(1-b)(\sqrt{2}y_0-\alpha_y)\tanh r_B}{b+\cosh(2r_B)} \\
    \mu_{q_B}^- &= \frac{[\sqrt{2}(1-b)x_0 - (1 + b + 2 \cosh 2r_B)\alpha_x] \tanh r_B}{b + \cosh 2r_B} \\
    \mu_{p_B}^- &= \frac{[\sqrt{2}(1-b)y_0 - (1 + b + 2 \cosh 2r_B)\alpha_y] \tanh r_B}{b + \cosh 2r_B}
\end{align}
More succinctly, after applying $\hat D(-g_\alpha\langle \bm \alpha \rangle)$ the non-Gaussian Wigner function of the total output state becomes a probabilistic mix of $\left[ W_{AB} \right]_{x_0,y_0}$, each of which is a Gaussian state represented by the same covariance matrix but different displacements $\bm{\mu}^{\pm\pm} = (\mu_{q_A}, \mu_{p_A}, \mu_{q_B}^\pm, \mu_{p_B}^\pm)^T$. Labelling these states as $\left[ W_{AB}^{d,\pm \pm} \right]_{x_0,y_0}$:
\begin{align}
    \left[ W_{AB}^d \right]_{x_0,y_0} \longrightarrow \begin{cases}
        \left[ W_{AB}^{d,++} \right]_{x_0,y_0} & x_0 \geq 0, y_0 \geq 0 \vspace{0.0cm} \\
        \left[ W_{AB}^{d,+-} \right]_{x_0,y_0} & x_0 \geq 0, y_0 < 0 \vspace{0.0cm} \\
        \left[ W_{AB}^{d,-+} \right]_{x_0,y_0} & x_0 < 0, y_0 \geq 0 \vspace{0.0cm} \\
        \left[ W_{AB}^{d,--} \right]_{x_0,y_0} & x_0 < 0, y_0 < 0.
    \end{cases}
\end{align}
We can now proceed to calculate mean and covariance of the output state after Bob has performed both postprocessing operations. Importantly, we must calculate the ensemble mean and covariance, since the state $\left[ W_{AB}^d \right]_{x_0,y_0}$ is necessarily dependent on the measurement outcomes, which vary shot-by-shot. To do this, we can exploit the fact that each shot of the protocol is an independent and identically distributed state represented by $\left[ W_{AB}^d \right]_{x_0,y_0}$ and distributed according to $P(x_0,y_0)$; the ensemble state is thus the continuous mixture
\begin{align}
    W_{AB}^d &= \int \dd x_0 \ \dd y_0 \ P(x_0,y_0) \left[ W_{AB}^d \right]_{x_0,y_0}.
\end{align}
The moments of the state are now given by Eq. \eqref{eq:wignerfunc_moments}. For example:
\begin{align}
    \mu_{AB, \ \hat q_A}^d = &\Tr \left[ \hat q_A \hat \rho_{AB}^d\right] \\
    = &\iint \dd^2\bm{x} \ \dd^4 \bm{q} \ P(x_0,y_0) \left[ W_{AB}^d \right]_{x_0,y_0} q_A \\
    = &\int_0^\infty \int_0^\infty \dd^2\bm{x} \int \dd^4 \bm{q} \ P(x_0,y_0) \left[ W_{AB}^{d,++} \right]_{x_0,y_0} q_A +\int_0^\infty \int_{-\infty}^0 \dd^2\bm{x} \int \dd^4 \bm{q} \ P(x_0,y_0) \left[ W_{AB}^{d,+-} \right]_{x_0,y_0} q_A \notag \\
    + &\int_{-\infty}^0 \int_0^\infty \dd^2\bm{x} \int \dd^4 \bm{q} \ P(x_0,y_0) \left[ W_{AB}^{d,-+} \right]_{x_0,y_0} q_A + \int_{-\infty}^0 \int_{-\infty}^0 \dd^2\bm{x} \int \dd^4 \bm{q} \ P(x_0,y_0) \left[ W_{AB}^{d,--} \right]_{x_0,y_0} q_A \\
    = &\int_0^\infty \int_0^\infty \dd^2\bm{x} \ P(x_0,y_0) \left[ \frac{c(\sqrt{2}x_0 - \alpha_x)}{b + \cosh 2r_B} \right] +\int_0^\infty \int_{-\infty}^0 \dd^2\bm{x} \ P(x_0,y_0) \left[ \frac{c(\sqrt{2}x_0 - \alpha_x)}{b + \cosh 2r_B} \right] \notag\\
    + &\int_{-\infty}^0 \int_0^\infty \dd^2\bm{x} \ P(x_0,y_0) \left[ \frac{c(\sqrt{2}x_0 - \alpha_x)}{b + \cosh 2r_B} \right] + \int_{-\infty}^0 \int_{-\infty}^0 \dd^2\bm{x} \ P(x_0,y_0) \left[ \frac{c(\sqrt{2}x_0 - \alpha_x)}{b + \cosh 2r_B} \right] \\
    = &\iint \dd^2\bm{x} \ P(x_0,y_0) \left[ \frac{c(\sqrt{2}x_0 - \alpha_x)}{b + \cosh 2r_B} \right] = 0 \\
    \mu_{AB, \ \hat q_B}^d = &\Tr \left[ \hat q_B \hat \rho_{AB}^d \right] \\
    = &\iint \dd^2\bm{x} \ \dd^4 \bm{q} \ P(x_0,y_0) \left[ W_{AB}^d \right]_{x_0,y_0} q_B \\
    = &\int_0^\infty \int_0^\infty \dd^2\bm{x} \int \dd^4 \bm{q} \ P(x_0,y_0) \left[ W_{AB}^{d,++} \right]_{x_0,y_0} q_B +\int_0^\infty \int_{-\infty}^0 \dd^2\bm{x} \int \dd^4 \bm{q} \ P(x_0,y_0) \left[ W_{AB}^{d,+-} \right]_{x_0,y_0} q_B \notag \\
    + &\int_{-\infty}^0 \int_0^\infty \dd^2\bm{x} \int \dd^4 \bm{q} \ P(x_0,y_0) \left[ W_{AB}^{d,-+} \right]_{x_0,y_0} q_B + \int_{-\infty}^0 \int_{-\infty}^0 \dd^2\bm{x} \int \dd^4 \bm{q} \ P(x_0,y_0) \left[ W_{AB}^{d,--} \right]_{x_0,y_0} q_B \\
    = &\int_0^\infty \int_0^\infty \dd^2\bm{x} \ P(x_0,y_0) \left[ \mu_{q_B}^+ \right] +\int_0^\infty \int_{-\infty}^0 \dd^2\bm{x} \ P(x_0,y_0) \left[ \mu_{q_B}^+ \right] \notag\\
    + &\int_{-\infty}^0 \int_0^\infty \dd^2\bm{x} \ P(x_0,y_0) \left[ \mu_{q_B}^- \right] + \int_{-\infty}^0 \int_{-\infty}^0 \dd^2\bm{x} \ P(x_0,y_0) \left[ \mu_{q_B}^- \right] \\
    = &\ -2 \alpha_x \tanh r_B e_C.
\end{align}
The remaining moments follow in the same way. Lastly, we extend to the case where Alice sends not just a single arbitrary displacement, but instead randomly selects a displacement from the classical alphabet described in the main work. For each shot, the sent state thus becomes a stochastic four-way mix of the state describing each possible symbol:
\begin{align}
    \hat \rho_{AA_S} \longrightarrow \frac{\hat \rho_{AA_S}^{d_1} + \hat \rho_{AA_S}^{d_2} + \hat \rho_{AA_S}^{d_3} + \hat \rho_{AA_S}^{d_4}}{4}.
\end{align}
Because the protocol is composed of entirely linear operations and measurements, the only effect is to transform the output state into a balanced mix of the four single-symbol states such that
\begin{align}
    \left[ W_{AB}^\mathrm{out} \right]_{x_0,y_0} \longrightarrow \frac{1}{4}\begin{cases}
        \left[ W_{AB}^{d_1,++} \right]_{x_0,y_0} + \left[ W_{AB}^{d_2,++} \right]_{x_0,y_0} + \left[ W_{AB}^{d_3,++} \right]_{x_0,y_0} + \left[ W_{AB}^{d_4,++} \right]_{x_0,y_0} & x_0 \geq 0, y_0 \geq 0 \vspace{0.25cm} \\
        \left[ W_{AB}^{d_1,+-} \right]_{x_0,y_0} + \left[ W_{AB}^{d_2,+-} \right]_{x_0,y_0} + \left[ W_{AB}^{d_3,+-} \right]_{x_0,y_0} + \left[ W_{AB}^{d_4,+-} \right]_{x_0,y_0}
        & x_0 \geq 0, y_0 < 0 \vspace{0.25cm} \\
        \left[ W_{AB}^{d_1,-+} \right]_{x_0,y_0} + \left[ W_{AB}^{d_2,-+} \right]_{x_0,y_0} + \left[ W_{AB}^{d_3,-+} \right]_{x_0,y_0} + \left[ W_{AB}^{d_4,-+} \right]_{x_0,y_0} & x_0 < 0, y_0 \geq 0 \vspace{0.25cm} \\
        \left[ W_{AB}^{d_1,--} \right]_{x_0,y_0} + \left[ W_{AB}^{d_2,--} \right]_{x_0,y_0} + \left[ W_{AB}^{d_3,--} \right]_{x_0,y_0} + \left[ W_{AB}^{d_4,--} \right]_{x_0,y_0} & x_0 < 0, y_0 < 0.
    \end{cases}
\end{align}
Calculation of the moments also proceeds straightforwardly in the many-symbol case, since expectation values are also linear. Thus
\begin{align}
    \mu_{AB, \ \hat q_A}^\mathrm{out} = \Tr \left[ \hat q_A \hat \rho_{AB}^\mathrm{out} \right] &= \frac{\mu_{AB, \ \hat q_B}^{d_1} + \mu_{AB, \ \hat q_B}^{d_2} + \mu_{AB, \ \hat q_B}^{d_3} + \mu_{AB, \ \hat q_B}^{d_4}}{4} \\
    &= 0 \\
    \mu_{AB, \ \hat q_B}^\mathrm{out} = \Tr \left[ \hat q_A \hat \rho_{AB}^\mathrm{out} \right] &= \frac{\mu_{AB, \ \hat q_B}^{d_1} + \mu_{AB, \ \hat q_B}^{d_2} + \mu_{AB, \ \hat q_B}^{d_3} + \mu_{AB, \ \hat q_B}^{d_4}}{4} \\
    &= \frac{1}{4} \left[ -2\alpha_x \tanh r_B e_C - 2\alpha_x \tanh r_B e_C + 2\alpha_x \tanh r_B e_C + 2 \alpha_x \tanh r_B e_C \right] \\
    &= 0
\end{align}
and similarly for the remaining moments of the state. Thus, the covariance matrix of the ensemble state $\hat \rho_{AB}^\mathrm{out}$ for the protocol with full classical communication is
\begin{align}
    \bm{\mu}_{AB}^\mathrm{out} &= (0,0,0,0)^T \\
    V_{AB}^\mathrm{out} &= \begin{pmatrix}
        a' \id & c' \sz \\
        c' \sz & b' \id
    \end{pmatrix}
\end{align}
for 
\begin{align}
    a' &= a \\
    b' &= b \tanh^2 r_B + \sech^2 r_B + 2(1-b) \tanh^2r_B \frac{\alpha e^{-\frac{1}{4}\frac{\alpha^2}{b + \cosh2r_B}}}{\sqrt{\pi}\sqrt{b + \cosh 2r_B}} + \alpha^2 \mathrm{erfc\left( \frac{\alpha}{2\sqrt{b + \cosh 2r_B}} \right) \tanh^2 r_B} \\
    c' &= -c \tanh r_B \left[ 1 - \frac{\alpha e^{-\frac{1}{4}\frac{\alpha^2}{b + \cosh2r_B}}}{\sqrt{\pi}\sqrt{b + \cosh2r_B}} \right].
\end{align}

\section{Estimating entanglement}
\label{app:ent}

\subsection{Reverse coherent information}
\label{app:ent:rci}
The reverse coherent information (RCI) is defined in the following way \cite{garcia-patron_reverse_2009}. Suppose the Gaussian CMQC protocol is represented by the quantum channel
\begin{align}
    \Lambda : \hat \rho_\mathrm{in} \rightarrow \hat \rho_{AB}^\mathrm{out}.
\end{align}
The maximum amount of entanglement distillable from the state $\hat \rho_{AB}^\mathrm{out}$ in e-bits is by construction identical to the entanglement distribution capacity of $\Lambda$, since we assume that the input state $\rho^\mathrm{in}$ is independent and identically distributed (recalling that each input state is a superposition of the four individual single-symbol states). Calculating the entanglement distribution capacity under arbitrary two-way communication $\mathcal{E}_\leftrightarrow$ is not technically feasible; a lower bound is the entanglement distribution capacity assisted by a simplified reverse communication $\mathcal{E}_\triangleleft$ \footnote{It may seem unreasonable to claim that Bob does not have a backwards communications channel, as we do in the body of the paper, and then quantify the distillable entanglement explicitly in terms of a reverse-assisted capacity. We are obliged to ignore this apparent contradiction for practical reasons, since the forward capacity is trivially known and not especially illustrative. However, practical implementations of the protocol would likely have two-way communication accessible in some form or another and so evaluating the RCI is not unreasonable.}, which is itself lower-bounded by the so-called reverse coherent information capacity
\begin{align}
    \mathcal{E}_R(\Lambda) \equiv \max_{\hat \rho^\mathrm{in}} S(\hat \rho_{A}^\mathrm{out}) - S(\hat \rho_{AB}^\mathrm{out})
\end{align}
for $\hat \rho_A^\mathrm{out} = \Tr_B[\hat \rho_{AB}^\mathrm{out}]$. The reverse coherent information $\mathcal{R}_\mathrm{out}$ of the state $\hat \rho_{AB}^\mathrm{out}$ for a fixed input, given by
\begin{align}
    \mathcal{R}_\mathrm{out} &= S(\hat \rho_{A}^\mathrm{out}) - S(\hat \rho_{AB}^\mathrm{out})
\end{align}
therefore bounds the distillable entanglement \cite{weedbrook_gaussian_2012, winnel_generalized_2020, pirandola_advances_2020}. Lastly, by Gaussian extremality the quantity $\mathcal{R}_\mathrm{out}$ is lower-bounded by the RCI of the zero-mean Gaussian state $\hat \rho_{AB}^G$ defined by the covariance matrix $V_{AB}^\mathrm{out}$, and so we calculate a lower bound $\mathcal{R}$ on the distillable entanglement as
\begin{align}
    \mathcal{R}_\mathrm{out} \geq \mathcal{R} &\equiv S(\hat \rho_{A}^G) - S(\hat \rho_{AB}^G).
\end{align}
The quantities $S(\hat \rho_{A}^G)$, $S(\hat \rho_{AB}^G)$ are calculated from the symplectic eigenvalues $\nu_i$ of the respective covariance matrices describing each state or substate \cite{weedbrook_gaussian_2012}:
\begin{align}
    S(\hat \rho^G_{AB}) &= g(\nu_{AB}^+) + g(\nu_{AB}^-), \\
    S(\hat \rho^G_{A}) &= g(\nu_A),
\end{align}
where
\begin{align}
    \nu_{AB}^\pm &= \sqrt{\frac{1}{2} \LL( D_1 \pm \sqrt{D_1^2 - 4 D_2^2} \RR)}, \\
    D_1 &= a^2 + b^2 - 2c^2, \\
    D_2 &= a b - c^2
\end{align}
and $\nu_A = a$. Here $g$ is the usual function \cite{weedbrook_gaussian_2012}
\begin{align}
    g(x) &\equiv \frac{x+1}{2} \log_2 \left( \frac{x+1}{2} \right) - \frac{x-1}{2} \log_2 \left( \frac{x-1}{2} \right).
\end{align}

\subsection{Entanglement of formation}
\label{app:ent:eof}
The entanglement of formation $\mathcal{E}_F$ describes the minimum amount of entanglement on average, in e-bits, required to produce an arbitrary entangled state \cite{bennett_mixed-state_1996}. It has an equivalent definition in the context of continuous-variable states as the `\textit{minimum amount of two-mode squeezing required to prepare an} (arbitrary) \textit{entangled state}' \cite{tserkis_quantifying_2017}. Estimating $\mathcal{E}_F$ for a given state is a hard problem requiring optimisation over a convex roof \cite{tserkis_quantifying_2019}, and for the majority of cases no analytic solution exists. For the state $\hat \rho_{AB}^\mathrm{out}$ the best approximation is again the lower bound provided by Gaussian extremality \cite{wolf_extremality_2006}, we therefore calculate the quantity $\mathcal{E}_F$ such that
\begin{align}
    \mathcal{E}_F &\equiv \mathcal{E}_F(\hat \rho_{AB}^G) \leq \mathcal{E}_F(\hat \rho_{AB}^\mathrm{out}).
\end{align}
We define $\mathcal{E}_F(\hat \rho_{AB}^G)$ as the entanglement of formation for the Gaussian state with covariance matrix $V_{AB}^\mathrm{out}$ and compute it using the numerical approach outlined by Tserkis, Onoe and Ralph \cite{tserkis_quantifying_2019} for an arbitrary two-mode Gaussian state; letting
\begin{align}
    V_{AB}^\mathrm{out} &\equiv \begin{pmatrix}
        a & 0 & c_1 & 0 \\
        0 & a & 0 & c_2 \\
        c_1 & 0 & b & 0 \\
        0 & c_2 & 0 & b
    \end{pmatrix}
\end{align}
$\mathcal{E}_F$ is given by
\begin{align}
    \mathcal{E}_F &= \inf_{r_{-} \ \leq \ r \ \leq \ r_+} H\{k(r)\} = \cosh^2 \log_2(\cosh^2k) - \sinh^2 \log_2(\sinh^2k)
\end{align}
where
\begin{align}
    k(r) &= \frac{1}{2} \cosh^{-1} \left( e^{2r_1} \chi \cosh^2 r + e^{2r_2} \chi \sinh^2 r \right) \\
    \chi &= \sqrt{\frac{e^{-2r_1} + e^{-2r_2} \tanh^2 r}{e^{2r_1} + e^{2r_2} \tanh^2 r}}
\end{align}
with
\begin{align}
    r_1 &= \ln \sqrt{\frac{(a-b)\xi_+ - 2\theta\sinh2r - (a+b)\xi_- \cosh2r}{\omega - \sigma + 1 + \sqrt{\gamma (\zeta_1 + \zeta_2)}}} \\
    r_2 &= \ln \sqrt{\frac{(a-b)\xi_+ + 2\theta\sinh2r + (a+b)\xi_- \cosh2r}{\omega + \sigma - 1 + \sqrt{\gamma (\zeta_1 + \zeta_2)}}} \\
    \xi_\pm &= ab - c_1^2 \pm 1 \\
    \theta &= abc_2 - c_1^2c_2 + c_1 \\
    \omega &= (a - b)\left[ (a + b) \cosh2r + (c_2 - c_1)\sinh2r \right] \\
    \gamma &= \frac{1}{2} \left[ a^2 (b^2 - 1) - ab(c_1^2 + c_2^2) - b^2 + (c_1c_2 - 1)^2 \right] \\
    \zeta_1 &= a^2 (2b^2 - 1) - 2ab(c_1^2 + c_2^2 - 1) - b^2 + 2c_1^2c_2^2 + 2 \\
    \zeta_2 &= 2(a+b)(c_1 - c_2)\sinh4r - \left[ (a+b)^2 - 4c_1c_2 \right]\cosh4r \\
    \sigma &= \det V_{AB}^\mathrm{out}.
\end{align}
The bounds $r_\pm$ are given by \cite{tserkis_quantifying_2017}
\begin{align}
    r_\pm &= \frac{1}{2} \ln \sqrt{\frac{\kappa \pm \sqrt{\kappa^2 - \lambda_+ \lambda_-}}{\lambda_-}}
\end{align}
for
\begin{align}
    \kappa &= 2(\sigma + 1) - (a - b)^2 \\
    \lambda_\pm &= a^2 + b^2 - 2 c_1 c_2 + 2\left[ (ab - c_1c_2) \pm (c_1 - c_2)(a + b) \right].
\end{align}

\subsection{Secret-key rate calculations}
\label{app:ent:qkd}
In general, the asymptotic secret-key generation rate $\mathcal{K}$ for a quantum key distribution scheme using reverse reconciliation is given by the Devetak-Winter bound \cite{devetak_capacity_2005, laudenbach_continuousvariable_2018}
\begin{align}
    \mathcal{K} &= \beta I_{AB} - \chi_{EB}
\end{align}
for $\beta = 0.95$ the asymptotic error reconciliation efficiency, $I_{AB}$ the classical mutual information, and $\chi_{EB}$ the Holevo information between Bob and an attacker Eve. Again, calculating $I_{AB}$ or $\chi_{EB}$ analytically for $\hat \rho_{AB}^\mathrm{out}$ is a non-trivial task. We first exploit extremality \cite{wolf_extremality_2006} to produce a lower bound on the mutual information. Writing 
\begin{align}
    V_{AB}^\mathrm{out} &\equiv \begin{pmatrix}
        a\id & c\sz \\
        c\sz & b\id
    \end{pmatrix}
\end{align}
and assuming Alice and Bob use a no-switching (heterodyne) \cite{weedbrook_quantum_2004} measurement scheme, then
\begin{align}
    I_{AB}(\hat \rho_{AB}^\mathrm{out}) &\geq I_{AB}(\hat \rho_{AB}^G) = \log_2 \left( \frac{a + 1}{a + 1 -\frac{c^2}{b+1}} \right).
\end{align}
Similarly, the optimality of Gaussian attacks \cite{navascues_optimality_2006, garcia-patron_unconditional_2006} also upper-bounds the Holevo information by the equivalent Gaussian amount. Assuming that Eve holds a purification $\hat \rho_{ABE}$ of the Gaussian state $\hat \rho_{AB}^G$, then
\begin{align}
    \chi_{BE}(\hat \rho_{AB}^\mathrm{out}) \leq \chi_{BE}(\hat \rho_{AB}^G) &\equiv \chi_{BE}(\hat \rho_{ABE}) \\
    &= S(\hat \rho_E) - S(\hat \rho_{E|B}) \\
    &= S(\hat \rho_{AB}^G) - S(\hat \rho_{A|B}^G)
\end{align}
where
\begin{align}
    S(\hat \rho_{A|B}^G) &= g(\nu_{A|B}) = g\left( a - \frac{c^2}{b+1} \right).
\end{align}
Thus we compute a lower bound $\mathcal{K}^\infty$ on the asymptotic keyrate
\begin{align}
    \mathcal{K} \geq \mathcal{K}^\infty = \beta I_{AB}(\hat \rho_{AB}^G) - \chi_{EB}(\hat \rho_{AB}^G).
\end{align}

\end{widetext}

\end{document}